\documentclass[12pt,fleqn]{article}
\usepackage{graphicx,cite}

\def\theequation{\thesection.\arabic{equation}}
\newcommand{\newsection}[1]{\section{#1}\setcounter{equation}{0}}
\newcommand{\newappendix}[1]{\section*{#1}\setcounter{equation}{0}}

\def\be{\begin{equation}}
\def\ee{\end{equation}}
\def\bea{\begin{eqnarray}}
\def\eea{\end{eqnarray}}
\def\nnb{\nonumber}
\def\bbuildrel#1_#2^#3{\mathrel{\mathop{\kern 0pt#1}\limits_{#2}^{#3}}}
\def\slash#1{\setbox0=\hbox{$#1$}#1\hskip-\wd0\dimen0=5pt\advance
       \dimen0 by-\ht0\advance\dimen0 by\dp0\lower0.5\dimen0\hbox
         to\wd0{\hss\sl/\/\hss}}

\newcommand{\scs}{\scriptscriptstyle}
\newcommand{\f}{\frac}
\newcommand{\fm}[2]{{\textstyle \frac{#1}{#2}}}

\newcommand{\al}{\widetilde{\alpha}}
\newcommand{\ep}{\epsilon}
\newcommand{\psla}{p\!\!\!/\,}
\newcommand{\ksla}{k\!\!\!/\,}

\begin{document}

\begin{titlepage}

\begin{flushright}
ZU-TH 25/06\\
CERN-PH-TH/2006-271\\
IFT-23/2006\\
hep-ph/0612329\\[2cm]
\end{flushright}

\begin{center}
\setlength {\baselineskip}{0.3in} 
{\bf\Large Four-Loop Anomalous Dimensions for Radiative Flavour-Changing Decays}\\[2cm]
\setlength {\baselineskip}{0.2in}
{\large  Micha{\l} Czakon,$\!\!^{1,2}$~ Ulrich Haisch$^3$~ and~ Miko{\l}aj Misiak$^{4,5}$}\\[5mm]

$^1$~{\it Institut f\"ur Theoretische Physik und Astrophysik,
          Universit\"at W\"urzburg,\\
          Am Hubland, D-97074 W\"urzburg, Germany.}\\[5mm]
$^2$~{\it Department of Field Theory and Particle Physics, 
          Institute of Physics,\\
          University of Silesia, Uniwersytecka 4, PL-40007 Katowice, Poland.}\\[5mm]
$^3$~{\it Institut f\"ur Theoretische Physik, Universit\"at of Z\"urich,\\
          Winterthurerstrasse 190, CH-8057 Z\"urich, Switzerland.}\\[5mm]
$^4$~{\it Theoretical Physics Division, CERN, CH-1211 Geneva 23, Switzerland.}\\[5mm] 
$^5$~{\it Institute of Theoretical Physics, Warsaw University,\\
           Ho\.za 69, PL-00681 Warsaw, Poland.}\\[15mm] 
{\bf Abstract}\\[5mm]
\end{center} 
\setlength{\baselineskip}{0.2in} 

We evaluate the complete four-loop anomalous dimension matrix that is
necessary for\linebreak determining the effective flavour-changing
  neutral current couplings~ $\bar qq'\gamma$~ and~ $\bar qq'g$~ at the
next-to-next-to-leading order in QCD. The resulting~ ${\cal
  O}(\alpha_s^2(\mu_b))$~ correction to the~ ${\bar B}\to X_s\gamma$~
branching ratio amounts to around~ $-2.9$\%~ for $\mu_b=5\,$GeV, and~
$-4.4$\%~ for $\mu_b=2.5\,$GeV.

\end{titlepage}

\newsection{Introduction \label{sec:intro}}

Flavour-Changing Neutral Current (FCNC) decays are well known to provide
stringent constraints on extensions of the Standard Model (SM) because their
SM amplitudes are loop-suppressed. An additional chirality suppression factor
$m_q/M_W$ for radiative FCNC transitions $q \to q'\gamma$ makes these
processes particularly attractive. The chirality suppression may be off-set in
certain new-physics models like the Minimal Supersymmetric Standard Model
with large $\tan\beta$~\cite{Bertolini:1990if} or left-right models~\cite{Cho:1993zb}.
Precise SM calculations of the radiative flavour-changing decay rates are thus
vital for our ability to derive constraints on new physics from these
observables.

In the present paper, we evaluate the complete four-loop Anomalous
Dimension Matrix (ADM) that is necessary for calculating the effective
FCNC couplings $\bar qq'\gamma$ and $\bar qq'g$ at the
Next-to-Next-to-Leading Order (NNLO) in QCD.  Our main motivation is
contributing to the NNLO calculation of the inclusive ${\bar B}\to
X_s\gamma$ branching ratio that is generated at the quark level by the
$b \to s\gamma$ transition. For this decay, both the current
experimental errors and the non-perturbative effects are smaller than
the perturbative NNLO corrections \cite{Misiak:2006zs}. Our four-loop
${\cal O}(\alpha_s^2(\mu_b))$ contributions turn out to suppress
${\cal B}({\bar B}\to X_s\gamma)$ by around $2.9$\% for
$\mu_b=5\,$GeV, and 4.4\% for $\mu_b=2.5\,$GeV. The dominant part of
this effect has already been included in the phenomenological NNLO
analysis of Refs.~\cite{Misiak:2006zs,Misiak:2006ab}.

The article is organized as follows. In Section~\ref{sec:Leff}, we introduce
the relevant effective Lagrangian and describe the structure of the
Renormalization Group Equations (RGEs) for the Wilson coefficients. In
Section~\ref{sec:bare4loop}, the bare four-loop calculation is described.
Section~\ref{sec:subdiv} is devoted to the calculation of subdivergences. Our
final results for the renormalization constants and ADMs are given in
Section~\ref{sec:ADM}. The numerical size of the four-loop effects is analyzed in
Section~\ref{sec:numerics}. We summarize in Section~\ref{sec:summary}. 
Appendix~A contains solutions to the RGEs for all the Wilson coefficients
that matter for $b\to s \gamma$ at the NNLO before including higher-order
electroweak corrections.

\newsection{Effective Lagrangian \label{sec:Leff}}

All the FCNC processes that have been observed so far are generated at the
electroweak scale $\mu_0 \sim M_W, m_t$, while the relevant momentum transfer
scale $\mu_f$ is much lower. Large logarithm $\left( \alpha_s \ln
  \mu_0^2/\mu_f^2\right)^n$ resummation at each order of the perturbation
series in $\alpha_s(\mu_f)$ is necessary to obtain viable results. It is
achieved by making use of the RGEs in an effective theory that arises from the
SM after decoupling of the heavy electroweak bosons and the top
quark~\cite{Buchalla:1995vs}.

The effective Lagrangian that matters for the~ $q \to q'\gamma$~ and~
$q \to q'g$~ transitions at the leading order in the electroweak interactions
is given below. For definiteness, we specify the quark flavours as in the $b\to
s\gamma$ case.
\bea 
{\cal L}_{\rm eff} &=& {\cal L}_{\scs {\rm QCD} \times {\rm QED}}(u,d,s,c,b)~
+~ \f{4 G_F}{\sqrt{2}} V^*_{ts} V_{tb} \sum_{i=1}^{8} C_i(\mu) Q_i~ +~ \ldots.
\label{Leff2}
\eea
Here, $G_F$ and $V_{ij}$ stand for the Fermi coupling constant
and the quark mixing matrix elements, respectively. The Wilson
coefficients $C_i(\mu)$ play the role of coupling constants at the effective
weak vertices (operators) $Q_i$.  The operator basis is chosen as in
Ref.~\cite{Chetyrkin:1996vx} to avoid difficulties with $\gamma_5$ in
dimensional regularization.
\be \label{physical}
\begin{array}{rl}
Q_1 ~= & (\bar{s} \gamma_{\mu} T^a   P_L c)        (\bar{c} \gamma^{\mu} T^a P_L b),\\[2mm]
Q_2 ~= & (\bar{s} \gamma_{\mu}       P_L c)        (\bar{c} \gamma^{\mu}     P_L b),\\[2mm]
Q_3 ~= & (\bar{s} \gamma_{\mu}       P_L b) \sum_q (\bar{q} \gamma^{\mu}         q),\\[2mm]     
Q_4 ~= & (\bar{s} \gamma_{\mu} T^a   P_L b) \sum_q (\bar{q} \gamma^{\mu} T^a     q),\\[2mm]    
Q_5 ~= & (\bar{s} \gamma_{\mu_1}
                  \gamma_{\mu_2}
                  \gamma_{\mu_3}     P_L b) \sum_q (\bar{q} \gamma^{\mu_1} 
                                                            \gamma^{\mu_2}
                                                            \gamma^{\mu_3}       q),\\[2mm]     
Q_6 ~= & (\bar{s} \gamma_{\mu_1}
                  \gamma_{\mu_2}
                  \gamma_{\mu_3} T^a P_L b) \sum_q (\bar{q} \gamma^{\mu_1} 
                                                            \gamma^{\mu_2}
                                                            \gamma^{\mu_3} T^a   q),\\[2mm]
Q_7  ~= &  \f{e}{16\pi^2} \left[ \bar{s} \sigma^{\mu \nu} 
           ( m_s P_L + m_b P_R) b \right] F_{\mu \nu},\\[2mm]
Q_8  ~= &  \f{g}{16\pi^2} \left[ \bar{s} \sigma^{\mu \nu} 
           ( m_s P_L + m_b P_R) T^a b \right] G_{\mu \nu}^a,
\end{array}
\ee
where $P_{L,R} = (1\mp\gamma_5)/2$.  For simplicity, terms proportional to the
small $V_{ub}$ mixing have been neglected here. The dots on the 
r.h.s. of Eq.~(\ref{Leff2}) stand for ultraviolet (UV)
counterterms. Apart from $Q_1$,\ldots,$Q_8$, the counterterms contain
unphysical operators that matter only off-shell and/or in $D\neq 4$ spacetime
dimensions (see Section~\ref{sec:subdiv}).

Instead of the original $C_i(\mu)$, it is more convenient to work with certain
linear combinations of them, the so-called ``effective'' Wilson coefficients
\be \label{ceff}
C_i^{\rm eff}(\mu) = \left\{ \begin{array}{ll}
C_i(\mu), & \mbox{ for $i = 1, ..., 6$,} \\[1mm] 
C_7(\mu) + \sum_{j=1}^6 y_j C_j(\mu), & \mbox{ for $i = 7$,} \\[1mm]
C_8(\mu) + \sum_{j=1}^6 z_j C_j(\mu), & \mbox{ for $i = 8$.}
\end{array} \right.
\ee
The numbers $y_j$ and $z_j$ are defined so that the $b \to s \gamma$ and $b
\to sg$ amplitudes at the Leading Order (LO) in QCD are proportional to
$C_7^{\rm eff}$ and $C_8^{\rm eff}$, respectively~\cite{Buras:1994xp}.  In the
$\overline{\rm MS}$ scheme with fully anticommuting $\gamma_5$ which
is used in our calculation, one finds,~ $\vec{y} = (0, 0, -\f{1}{3},
-\f{4}{9}, -\f{20}{3}, -\f{80}{9})$ and $\vec{z} = (0, 0, 1, -\f{1}{6}, 20,
-\f{10}{3})$~\cite{Chetyrkin:1996vx}.\\[-3.5mm]

The RGEs for $C_i^{\rm eff}(\mu)$
\be \label{RGEs}
\mu \f{d}{d \mu} C_i^{\rm eff}(\mu) = C_j^{\rm eff}(\mu) \gamma^{\rm eff}_{ji}(\mu)
\ee
are governed by the ADM
\be 
\hat{\gamma}^{\rm eff}(\mu) = \sum_{n \geq 0} \al^{n+1} \hat{\gamma}^{(n)\rm eff}  
\ee
where $\al = \alpha_s(\mu)/(4\pi)$. This matrix is determined from
the effective theory renormalization constants.  The matrices
$\hat{\gamma}^{(n)\rm eff}$ have a block-triangular structure
\be \label{blocks}
\hat{\gamma}^{(n)\rm eff} = \left( \begin{array}{cc}
A^{(n)}_{6\times 6} &  B^{(n)}_{6\times 2} \\  
0_{2\times 6} &  C^{(n)}_{2\times 2} \end{array} \right).
\ee
The down-left block vanishes because the dipole operators $Q_7$ and
$Q_8$ are actually dimension-five ones, and thus generate no
UV divergences in dimension-six four-quark amplitudes. The
diagonal blocks $A^{(n)}$ and $C^{(n)}$ are
found from $(n+1)$-loop renormalization constants. Non-vanishing
contributions to the off-diagonal blocks $B^{(n)}$ arise
at two and more loops only. Once the normalization conventions for
$Q_7$ and $Q_8$ are chosen as in Eq.~(\ref{physical}), the blocks
$B^{(n)}$ contain information on $(n+2)$-loop renormalization.\footnote{
Inverse powers of coupling constants are sometimes used in the dipole
operator normalization to make $\hat{\gamma}^{(n)\rm eff}$ a purely
$(n+1)$-loop object. Such a choice is convenient at intermediate steps
(see Section~\ref{sec:ADM}), but the final results for the RGEs are
more compact when the normalization as in Eq.~(\ref{physical}) is
applied.}

The complete matrices $\hat{\gamma}^{(0)\rm eff}$ and
$\hat{\gamma}^{(1)\rm eff}$ together with the relevant references can
be found in Ref.~\cite{Chetyrkin:1996vx}.  We quote these results in
Section~\ref{sec:ADM}. The three-loop block $B^{(1)}$ was confirmed
in Ref.~\cite{Gambino:2003zm}.  At the NNLO, one needs to know the
full $8\times 8$ matrix $\hat{\gamma}^{(2)\rm eff}$. The three-loop
blocks $A^{(2)}$ and $C^{(2)}$ were calculated in
Refs.~\cite{Gorbahn:2004my} and \cite{Gorbahn:2005sa}, respectively.
In the present paper, we evaluate the four-loop block $B^{(2)}$.

\newsection{Bare Four-Loop Calculation \label{sec:bare4loop}}

\begin{figure}[t]
\begin{center}
\includegraphics[width=6cm,angle=0]{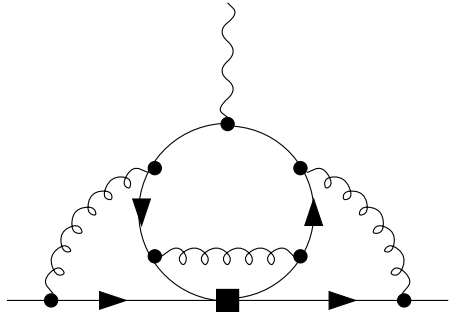}
\end{center}
\begin{center}
\caption{\sf One of the ${\cal O}(10^4)$ calculated four-loop diagrams. \label{fig:sample4loop}}
\end{center}
\end{figure}

The renormalization constants contributing to the matrix $B^{(2)}$ are
found after subloop renormalization from the UV-divergent parts of
four-loop diagrams like the one shown  in
Fig.~\ref{fig:sample4loop}. Each of the operators $Q_1$, \ldots $Q_6$
must be considered in the effective four-quark vertex, and the
external gauge boson can be either a photon or a gluon. The overall
number of relevant four-loop diagrams turns out to be 21986, when
different color and gamma matrix structures are treated together.

In order to simplify the calculation as much as possible, an infrared
  (IR) rearrangement as in\linebreak
Refs.~\cite{Misiak:1994zw,Chetyrkin:1997fm,vanRitbergen:1997va,Czakon:2004bu}
has been applied. In this approach, a Taylor expansion in the external
momenta is performed after introducing a common mass in all the
propagator denominators. The order of the expansion is determined by
the dimensions of the operator insertion and the Green's function.  In the
present case, up to two derivatives have to be applied. Afterwards,
the problem is reduced to evaluation of single-scale fully massive vacuum
integrals.

In principle, the contribution of any graph can be mapped onto the
following Dirac structures
\bea
  S_j &=& \left( \gamma_{\mu} \psla \ksla, \; 
    \gamma_{\mu} \; (p \cdot k), \;
    \gamma_{\mu} p^2, \;
    \gamma_{\mu} k^2, \;
    \psla k_{\mu}, \;
    \psla p_{\mu}, \;
    \ksla p_{\mu}, \;
    \ksla k_{\mu}, \;
  \right. \nonumber\\ && \hspace{4cm} \left. \label{DirStructures} 
    m_b \ksla \gamma_{\mu}, \;
    m_b \gamma_{\mu} \ksla, \;
    m_b \psla \gamma_{\mu}, \;
    m_b \gamma_{\mu} \psla, \;
    M^2 \gamma_{\mu} \right)_j,
\eea
where $p$ and $k$ stand for the incoming $b$-quark and the outgoing
photon/gluon momenta, respectively, whereas $M$ denotes the regulator
mass.  However, due to the huge computational resource requirements, only the
coefficients at the structures $S_j$ for $j=7,9,10,11$ have been calculated.
Only these four structures are needed to determine the sought ADM
$B^{(2)}$ (see Section~\ref{sec:subdiv}).

Comparing the vertex diagram structure and the Taylor expansion
  depth with those needed in the evaluation of the four-loop
$\beta$-function with a single gauge parameter, we note that the maximal
powers of numerators and denominators in the integrals are the same. This
has allowed us to use the solution of the integration-by-parts identities and
the master integrals determined in Ref.~\cite{Czakon:2004bu}. All the salient
details of the employed techniques can be found in that paper.

\newsection{Subdivergences \label{sec:subdiv}}

Considering renormalization of all the operators up to dimension six that can
arise in the effective theory (\ref{Leff2}) is necessary for proper
subtraction of subdivergences in our calculation. Since the four-loop diagrams
are evaluated in the usual Feynman-'t~Hooft gauge (without background fields),
in $D=4-2\ep$ dimensions and off-shell, the following types of
counterterms must be taken into account apart from the usual QCD ones:
\begin{itemize}
\item{} the original (``physical'') operators from Eq.~(\ref{physical}),
\item{} gauge-invariant operators that vanish by the QCD$\times$QED
        Equations Of Motion (EOM),
\item{} gauge-variant EOM-vanishing operators,
\item{} BRST-exact operators, i.e. operators that can be written as
        Becchi-Rouet-Stora-Tyutin variations of other operators, and
\item{} evanescent operators, i.e. operators that vanish in $D=4$
        spacetime dimensions by Dirac-algebra identities, but have to
        be included for $D\neq 4$.
\end{itemize}
One begins with writing down all the possible operators of dimension up to six
with proper flavour quantum numbers that fall into the above five classes. The
potentially infinite set of evanescent objects is reduced to a finite one at
each given order of the perturbation series once the prescription of
Ref.~\cite{Buras:1989xd} is applied.

The long list of relevant operators can be further reduced by taking into
account that the SM Lagrangian is invariant under the following discrete
transformations of its fields and parameters (already after the electroweak
symmetry breaking and mass matrix diagonalization):
\begin{itemize}
\item{} $b \leftrightarrow s$,~ 
        $m_b \leftrightarrow m_s$,~ 
        $V_{Qb} \leftrightarrow V_{Qs}$~ ($Q=u,c,t$).
\item{} $CP$ transformation and $V \to V^*$.
\end{itemize}
A combination of these two transformations maps the $\Delta B = -\Delta S = 1$
interaction terms onto themselves, so it must leave ${\cal L}_{\rm
  eff}$~(\ref{Leff2}) invariant. The relative plus sign between the $m_s$ and
$m_b$ terms in the dipole operators $Q_7$ and $Q_8$ is fixed by this very
symmetry. Once the operators are determined, terms proportional to $m_s$ are
often neglected.  Their effect on the $\bar{B} \to X_s \gamma$ branching ratio
is negligible (${\cal O}(m_s^2/m_b^2)$) but they matter for the CP asymmetry.
Some details on selecting the right number of gauge-invariant operators
have been given in Section~5.3 of Ref.~\cite{Bobeth:1999mk}.

Explicit expressions for all the operators that satisfy the above-mentioned
conditions can be found in Refs.~\cite{Gambino:2003zm,Gorbahn:2004my}, so we
do not repeat them here.  Let us only note that Section~2 of
Ref.~\cite{Gambino:2003zm} contains all the EOM-vanishing operators, including
the ones suppressed by the QED coupling $e$, and the one that gets generated
at three loops only ($Q_{22}$).  On the other hand, the only possible
BRST-exact operator and all the evanescent operators that we need here are
given in Section~3 of Ref.~\cite{Gorbahn:2004my}.  The BRST-exact operator
remains irrelevant. In particular, we have checked that the three-loop
analogue of the two-loop amplitude described in Fig.~3 of
Ref.~\cite{Gorbahn:2004my} turns out to mysteriously vanish, as well.

Apart from all the ``real'' off-shell counterterms that have been
mentioned so far, we encounter certain spurious ones that arise due to
our particular IR rearrangement.  The dimension-three gluon-mass-like
counterterm was discussed in detail in
Refs.~\cite{Misiak:1994zw,Chetyrkin:1997fm}. Spurious dimension-four
flavour-changing operators that are proportional to the IR-regulator
mass (see Eq.~(18) of Ref.~\cite{Gambino:2003zm}) are included, too.

The completeness of the full operators basis has been verified in many ways.
For instance, since the number of operators contributing at the tree level to
the $b \to s \gamma$ transition is smaller than the number of independent
Dirac structures $S_j$ (\ref{DirStructures}), there exist seven linear
relations between coefficients at these structures in the renormalized
amplitude. All these relations together with three analogous ones for
$b \to s g$ have been successfully tested up to three
loops~\cite{Chetyrkin:1996vx,Gambino:2003zm,Gorbahn:2004my,Misiak:2004ew}.  In
the present calculation, we determine coefficients at four of the structures
$S_j$ only, which leaves us with just one linear relation for $b \to s \gamma$
and none for $b \to s g$. Therefore, we essentially rely on the fact that the
operator basis is indeed complete.

Once the $(n\leq 3)$-loop diagrams with counterterms are added to the
bare four-loop ones, we obtain an expression of the form
\be \label{four.struct}
A_7 \ksla p_{\mu} + m_b \left( 
A_9 \ksla \gamma_{\mu} + 
A_{10} \gamma_{\mu} \ksla +
A_{11} \psla \gamma_{\mu} \right)
\ee
where the coefficients $A_i$ contain UV-poles only. These poles need to be
canceled by tree-level counterterms. First, we subtract such tree-level
counterterms that originate from products of the usual QCD renormalization
constants with the $(n\leq 3)$-loop operator ones, which modifies the
coefficients in Eq.~(\ref{four.struct}). Let us call the new coefficients
$A'_i$. The linear combination
\be \label{projection}
\f{1}{4} A'_7 - \f{1}{2} A'_9 + \f{1}{2} A'_{10} - \f{1}{2} A'_{11}
\ee
gives us the sought four-loop contributions to the renormalization constants
$Z_{i7}$ and $Z_{i8}$ (for $i=1,\ldots 6$) in the photonic and gluonic cases,
respectively. The single consistency relation that is successfully tested in
the photonic case reads $A'_9 + A'_{10} + A'_{11} = 0$. Proving that the
projection (\ref{projection}) is the right one requires studying Feynman rules
for each operator in the full basis and verifying that only $Q_7$ or $Q_8$ can
contribute to this particular combination at the tree level.

A very powerful test of the results in Eq.~(\ref{four.struct}) is provided by
the so-called locality constraints. Let us denote the $n$-loop~ 
$1/\ep^m$-contribution to $A_i$ by $A_i^{(nm)}$, i.e.
\be
A_i = \sum_{n,m=1}^4 \f{\mu^{2n\ep}}{\ep^m} A_i^{(nm)}.
\ee
The quantities $A_i^{(nm)}$ must satisfy many relations to ensure that
all the $(\ln^k\mu)/\ep^j$ poles cancel out in $A_i$. Such a
cancellation is necessary for the counterterms to remain polynomial
in external momenta, i.e. local in the position space. The explicit
relations read
\bea
12 A_i^{(44)} &=& -3 A_i^{(14)} ~=~ 2 A_i^{(24)} ~=~ -3 A_i^{(34)}\\
6 A_i^{(43)} &=&   -A_i^{(23)} - 3 A_i^{(33)}\\
 8 A_i^{(43)} &=&    A_i^{(13)} - 3 A_i^{(33)}\\ 
 4 A_i^{(42)} &=&   -A_i^{(12)} - 2 A_i^{(22)} - 3 A_i^{(32)}. 
\eea
All these relations have been checked to hold in the photonic and
gluonic cases for each of the six operator insertions. One should
realize that testing the $1/\ep^2$ poles in a four-loop amplitude
constitutes a much more sensitive cross-check of the calculation than
doing the same in a two-loop amplitude. Since $A_i^{4m}$ and
$A_i^{1m}$ have been evaluated with the help of different routines
written by different authors, we can be practically sure that our
calculation is free of errors like missing a diagram or taking it with
a wrong overall factor.

\newsection{Renormalization Constants and ADMs \label{sec:ADM}}

The counterterms in the effective Lagrangian that we have calculated
can be written in the following form
\be
\Delta {\cal L}^{\rm count}_{\rm eff} = \f{4 G_F}{\sqrt{2}} V^*_{ts} V_{tb} 
\sum_{i=1}^6 C_i \left( Z_{i7} X_7 + Z_{i8} X_8 \right),
\ee
where
\be \label{x7x8}
X_7 = \f{16\pi^2}{(Z_g g)^2} Z_0 Q_7 \hspace{2cm} \mbox{and} \hspace{2cm}
X_8 = \f{16\pi^2}{(Z_g g)^2} Z_0 Z_g Z_G^{1/2} Q_8.
\ee
Here, $Z_0$, $Z_g$ and $Z_G$ are the usual QCD renormalization constants
of the mass term, gauge coupling and the gluon field 
\be
(m\bar{\psi}\psi)_B = Z_0 m\bar{\psi}\psi, \hspace{2cm} 
g_B = Z_g g, \hspace{2cm} 
G^{a\,\mu}_B = Z_G^{1/2} G^{a\,\mu}. 
\ee
Below, we give our results for $Z_{i7}$ and $Z_{i8}$ up to four loops.
The ``trivial'' terms proportional to $\ln(4\pi)$ or to the Euler
constant $\gamma_{\scs E}$ are omitted, as they have no effect on the
$\overline{\rm MS}$ anomalous dimensions. The Riemann zeta function
value $\zeta(3) \simeq 1.20206$ is denoted by $\zeta_3$.  The number
of active flavours in the effective theory and the sum of their charges
are denoted by $f$ and $\overline{Q}$, respectively. For $Z_{17}$ and
$Z_{27}$, the charge $Q_u$ of the charm quark is retained
arbitrary. The incoming quark charge is set to $-\f{1}{3}$ in all the
expressions for $Z_{i7}$, which means that they correspond to the
down-type quark $\bar qq'\gamma$ coupling after substituting
$Q_u=\f{2}{3}$. For the external up-type quark case, one should
multiply all the terms in $Z_{i7}$ that come with no charge factor by
$-2$, and then replace $Q_u$ by $Q_d=-\f{1}{3}$. The same rule applies
to the matrices $B^{(n)}$ that are given in
Eqs.~(\ref{B0})--(\ref{B2}) at the end of this section. We find
\mathindent0cm
\bea
Z_{17} &=& 
-\fm{\al^2}{\ep} \left(\fm{4}{243} + \fm{1}{3} Q_u\right) 
+\fm{\al^3}{\ep} \left\{
\fm{1}{\ep} \left[ -\fm{274}{6561} -\fm{40}{2187} f
                   +\left( \fm{130}{27} - \fm{8}{27} f\right) Q_u \right]
+\fm{3239}{6561} -\fm{32}{6561} f - \left( \fm{352}{81} 
\right. \right. \nnb\\ &-& \left. \left. 
\fm{1}{81} f \right) Q_u \right\}
+\fm{\al^4}{\ep} \left\{
\fm{1}{\ep^2} \left[ \fm{97508}{59049} +\fm{487}{6561} f -\fm{104}{6561} f^2 
    -\left( \fm{1556}{27} -\fm{193}{27} f +\fm{2}{9} f^2 \right) Q_u -\fm{5}{108} \overline{Q} \right]
\right. \nnb\\ &+& \left. 
\fm{1}{\ep} \left[ -\fm{93353}{19683} +\fm{94513}{236196} f -\fm{100}{19683} f^2
    +\left( \fm{47807}{648} -\fm{599}{81} f +\fm{1}{81} f^2 \right) Q_u -\fm{245}{972} \overline{Q} \right]
-\fm{1385935}{8503056} -\fm{20435}{6561} \zeta_3 
\right. \nnb\\ &+& \left. 
\left( \fm{46625}{354294} + \fm{91}{729} \zeta_3 \right) f +\fm{70}{19683} f^2
+\left[ \fm{14797}{432} +\fm{8381}{162} \zeta_3 +\left( \fm{11969}{5832} 
      - \fm{8}{81} \zeta_3 \right) f +\fm{79}{1458} f^2 \right] Q_u
\right. \nnb\\ &+& \left. 
\left( \fm{3695}{3888} - \fm{25}{54} \zeta_3 \right) \overline{Q} \right\},\nnb\\[2mm]
Z_{18} &=& \fm{167}{648} \fm{\al^2}{\ep} 
+ \fm{\al^3}{\ep} \left[
\fm{1}{\ep} \left( -\fm{6098}{2187} + \fm{1373}{5832} f \right)
-\fm{40655}{34992} + \fm{431}{34992} f \right]
+ \fm{\al^4}{\ep} \left[
\fm{1}{\ep^2} \left(  \fm{11512021}{314928} 
                     -\fm{313309}{52488} f + \fm{839}{3888} f^2 \right) 
\right. \nnb\\ &+& \left. 
\fm{1}{\ep} \left( \fm{7084465}{839808} + \fm{575321}{314928} f + \fm{659}{34992} f^2 \right) 
-\fm{490821325}{22674816} -\fm{368107}{8748} \zeta_3 
- \left( \fm{2647291}{944784} + \fm{10381}{3888} \zeta_3 \right) f
-\fm{1955}{52488} f^2 \right],\nnb\\[2mm]
Z_{27} &=& \fm{\al^2}{\ep} \left( \fm{8}{81} + 2 Q_u \right) 
+\fm{\al^3}{\ep} \left\{
  \fm{1}{\ep} \left[ -\fm{4636}{2187} +\fm{80}{729} f 
                     -\left(\fm{206}{9} - \fm{16}{9} f \right) Q_u \right]
  +\fm{6698}{2187} +\fm{64}{2187} f + \left( \fm{128}{27} 
\right. \right. \nnb\\ &-& \left. \left. 
\fm{2}{27} f \right) Q_u \right\}
+\fm{\al^4}{\ep} \left\{
 \fm{1}{\ep^2} \left[ \fm{560390}{19683} -\fm{7400}{2187} f +\fm{208}{2187} f^2
                      + \left( \fm{4379}{18} -\fm{323}{9} f +\fm{4}{3} f^2 \right) Q_u
                       + \fm{5}{18} \overline{Q} \right]
\right. \nnb\\ &+& \left. 
\fm{1}{\ep} \left[ -\fm{735973}{13122} +\fm{173111}{39366} f +\fm{200}{6561} f^2
                     - \left( \fm{39427}{216} -\fm{613}{27} f +\fm{2}{27} f^2 \right) Q_u
                     + \fm{245}{162} \overline{Q} \right]
 +\fm{3142663}{1417176} +\fm{5068}{2187} \zeta_3 
\right. \nnb\\ &-& \left. 
 \left( \fm{79754}{59049} + \fm{146}{243} \zeta_3 \right) f 
 -\fm{140}{6561} f^2
 +\left[ \fm{11699}{36} -\fm{6140}{27} \zeta_3 
         - \left( \fm{17153}{972} + \fm{128}{27} \zeta_3 \right) f - \fm{79}{243} f^2 \right] Q_u
\right. \nnb\\ &+& \left. 
 \left( -\fm{3695}{648} + \fm{25}{9} \zeta_3 \right) \overline{Q} \right\},\nnb\\[2mm]
Z_{28} &=& \fm{19}{27} \fm{\al^2}{\ep} 
+ \fm{\al^3}{\ep} \left[ 
\fm{1}{\ep} \left( -\fm{23063}{2916} + \fm{571}{972} f \right) 
+\fm{5167}{2916} -\fm{917}{5832} f \right]
+ \fm{\al^4}{\ep} \left[
\fm{1}{\ep^2} \left( \fm{10632025}{104976} -\fm{63787}{4374} f +\fm{161}{324} f^2 \right) 
\right. \nnb\\ &-& \left. 
\fm{1}{\ep} \left(\fm{11481667}{139968} - \fm{714847}{52488} f +\fm{1307}{5832} f^2\right) 
+ \fm{252836197}{3779136} -\fm{487691}{5832} \zeta_3 
- \left(\fm{6112793}{629856} + \fm{1679}{648} \zeta_3\right) f
-\fm{2489}{17496} f^2 \right],\nnb\\[2mm]
Z_{37} &=& \fm{16}{81} \fm{\al^2}{\ep} 
+ \fm{\al^3}{\ep} \left[
\fm{1}{\ep} \left(-\fm{18236}{2187} +\fm{160}{729} f + 4 \overline{Q} \right)
+\fm{49144}{2187} -\fm{1816}{2187} f -\fm{56}{9} \overline{Q} \right]
+ \fm{\al^4}{\ep} \left\{
\fm{1}{\ep^2} \left[ \fm{2436463}{19683} -\fm{27850}{2187} f 
\right. \right. \nnb\\ &+& \left. \left.
\fm{416}{2187} f^2 
- \left( \fm{1265}{18} - \fm{14}{3} f \right) \overline{Q} \right]
+ \fm{1}{\ep} \left[-\fm{10574227}{26244} +\fm{1737559}{39366} f -\fm{5864}{6561} f^2
+ \left(\fm{76669}{648}-\fm{22}{3} f \right) \overline{Q}\right]
\right. \nnb\\ &+& \left. 
\fm{141396541}{708588} 
- \fm{59074}{2187} \zeta_3 
- \left( \fm{13789403}{472392} +\fm{598}{243} \zeta_3 \right) f
- \fm{244}{6561} f^2
+ \left[ \fm{18007}{324} - \fm{56}{3} \zeta_3 
-\left( \fm{1}{27} +\fm{8}{9} \zeta_3 \right) f \right] \overline{Q} 
\right\},\nnb\\[2mm]
Z_{38} &=& \fm{92}{27} \fm{\al^2}{\ep} 
+ \fm{\al^3}{\ep} \left[ 
\fm{1}{\ep} \left( -\fm{63293}{1458} +\fm{4085}{972} f \right) 
+\fm{87031}{1458} -\fm{3361}{1458} f \right] 
+\fm{\al^4}{\ep} \left[
\fm{1}{\ep^2} \left( \fm{32831857}{52488} -\fm{3821329}{34992} f +\fm{2891}{648} f^2 \right) 
\right. \nnb\\ &+& \left. 
\fm{1}{\ep} \left( -\fm{97744615}{69984} +\fm{73524781}{419904} f -\fm{19453}{5832} f^2 \right) 
+\fm{176999365}{118098} -\fm{1095749}{2916} \zeta_3 
- \left(\fm{248941103}{5038848} + \fm{505}{162} \zeta_3 \right) f
\right. \nnb\\ &+& \left. 
\left( \fm{27313}{34992} - \fm{5}{18} \zeta_3 \right) f^2 \right],\nnb\\[2mm]
Z_{47} &=& -\fm{\al^2}{\ep} \left(\fm{170}{243} -\fm{8}{81} f \right) 
+ \fm{\al^3}{\ep} \left[
  \fm{1}{\ep} \left( \fm{59824}{6561} -\fm{148}{81} f
                    +\fm{80}{729} f^2 +\fm{5}{3} \overline{Q} \right)
  -\fm{104951}{6561} + \fm{2774}{6561} f + \fm{16}{2187} f^2 -\fm{70}{27} \overline{Q} \right]
\nnb\\ &+& 
\fm{\al^4}{\ep} \left\{
    \fm{1}{\ep^2} \left[ -\fm{6446822}{59049} +\fm{521993}{19683} f
       -\fm{17827}{6561} f^2 +\fm{208}{2187} f^3
       - \left( \fm{850}{27} -\fm{20}{9} f \right) \overline{Q} \right]
  + \fm{1}{\ep} \left[ \fm{12186259}{39366} -\fm{7760143}{236196} f 
\right. \right. \nnb\\ &+& \left. \left.
       \fm{58249}{39366} f^2 +\fm{56}{6561} f^3
       + \left( \fm{206399}{3888} -\fm{1175}{324} f \right) \overline{Q} \right]
  - \fm{895573151}{2125764} - \fm{439534}{6561} \zeta_3 
  +\left( \fm{21721465}{708588} -\fm{8165}{2187} \zeta_3 \right) f
\right. \nnb\\ &-& \left. 
  \left( \fm{71315}{59049} +\fm{160}{243} \zeta_3 \right) f^2 -\fm{148}{6561} f^3
  - \left[ \fm{42823}{7776} - \fm{1355}{108} \zeta_3 
           -\left(\fm{305}{216} - \fm{25}{27} \zeta_3 \right) f \right] \overline{Q} \right\},\nnb\\[2mm]
Z_{48} &=& -\fm{\al^2}{\ep} \left( \fm{427}{324} +\fm{37}{216} f \right) 
+\fm{\al^3}{\ep} \left[
\fm{1}{\ep} \left(\fm{84661}{4374}+\fm{71}{36} f -\fm{185}{972} f^2 \right)
-\fm{540209}{17496} -\fm{40091}{34992} f -\fm{253}{2916} f^2 \right]
\nnb\\ &+& 
\fm{\al^4}{\ep} \left[
\fm{1}{\ep^2} \left( -\fm{44723201}{157464} -\fm{590221}{34992} f
                     +\fm{216085}{34992} f^2 -\fm{259}{1296} f^3 \right)
+ \fm{1}{\ep} \left( \fm{238543801}{419904} -\fm{4684067}{314928} f
                  -\fm{1281229}{419904} f^2 
\right. \right. \nnb\\ &-& \left. \left.
\fm{253}{1944} f^3 \right) -\fm{5633994047}{22674816} +\fm{910901}{4374} \zeta_3 
+\left(\fm{566791855}{30233088} + \fm{208123}{5832} \zeta_3\right) f
+ \left(\fm{2176469}{1259712} + \fm{115}{81} \zeta_3\right) f^2
+\fm{505}{34992} f^3 \right],\nnb\\[2mm]
Z_{57} &=& -\fm{464}{81} \fm{\al^2}{\ep} 
+ \fm{\al^3}{\ep} \left[
\fm{1}{\ep} \left( -\fm{118976}{2187} -\fm{10112}{729} f + 40 \overline{Q} \right)
+\fm{257344}{729} +\fm{16832}{2187} f -\fm{2288}{9} \overline{Q} \right]
+ \fm{\al^4}{\ep} \left\{
\fm{1}{\ep^2} \left[ \fm{28293352}{19683} 
\right. \right. \nnb\\ &+& \left. \left.
\fm{201176}{2187} f -\fm{32008}{2187} f^2
- \left(\fm{6160}{9} -\fm{140}{3} f \right) \overline{Q} \right]
+ \fm{1}{\ep} \left[ -\fm{17496454}{2187} +\fm{3399452}{19683} f +\fm{83368}{6561} f^2
+ \left( \fm{296492}{81} 
\right. \right. \right. \nnb\\ &-& \left. \left. \left.
\fm{752}{3} f \right) \overline{Q} \right]
+ \fm{1744567561}{177147} + \fm{9416120}{2187} \zeta_3 
- \left(\fm{59440775}{118098} - \fm{4708}{243} \zeta_3\right) f
- \left(\fm{92}{81} - \fm{16}{27} \zeta_3\right) f^2
\right. \nnb\\ &-& \left. 
\left[ \fm{1095949}{648} +\fm{1210}{9} \zeta_3 
-\left(\fm{1934}{27} - \fm{80}{9} \zeta_3\right) f \right] \overline{Q} \right\},\nnb\\[2mm]
Z_{58} &=& \fm{\al^2}{\ep} \left( \fm{2192}{27} + 9 f \right) 
- \fm{\al^3}{\ep} \left[
\fm{1}{\ep} \left( \fm{713164}{729} +\fm{5494}{243} f -8 f^2 \right)
-\fm{452714}{243} -\fm{97043}{1458} f +\fm{7}{3} f^2 \right]
+\fm{\al^4}{\ep} \left[
\fm{1}{\ep^2} \left( \fm{87867110}{6561} 
\right. \right. \nnb\\ &-& \left. \left.
                                           \fm{1639918}{2187} f
                     -\fm{21299}{162} f^2 +\fm{43}{6} f^3 \right)
- \fm{1}{\ep} \left( \fm{25933753}{729} -\fm{115452017}{52488} f 
                     -\fm{684959}{2916} f^2 +\fm{28}{9} f^3 \right) 
\right. \nnb\\ &+& \left. 
\fm{11439938489}{236196} -\fm{11082938}{729} \zeta_3 
-\left( \fm{269786863}{629856} + \fm{288731}{162} \zeta_3 \right) f
-\left( \fm{86423}{648} +\fm{635}{9} \zeta_3 \right) f^2 -\fm{37}{54} f^3 \right],\nnb\\[2mm]
Z_{67} &=& -\fm{\al^2}{\ep} \left(\fm{3008}{243} - \fm{80}{81} f - 12 \overline{Q}\right) 
+ \fm{\al^3}{\ep} \left\{
\fm{1}{\ep} \left[ \fm{925216}{6561} -\fm{19720}{729} f +\fm{800}{729} f^2 
+ \left( -\fm{404}{3} +\fm{32}{3} f \right) \overline{Q} \right] -\fm{389636}{2187} 
\right. \nnb\\ &+& \left. 
\fm{154280}{6561} f +\fm{448}{2187} f^2 
- \left(\fm{1468}{27} + \fm{28}{9} f\right) \overline{Q} \right\}
+ \fm{\al^4}{\ep} \left\{
\fm{1}{\ep^2} \left[ -\fm{82789640}{59049} +\fm{8283332}{19683} f
                -\fm{234328}{6561} f^2 +\fm{2080}{2187} f^3
\right. \right. \nnb\\ &+& \left. \left.
\left( \fm{37826}{27} -\fm{1885}{9} f +8 f^2 \right) \overline{Q} \right]
+ \fm{1}{\ep} \left[ \fm{20027615}{6561} -\fm{39955153}{59049} f 
           +\fm{708617}{19683} f^2 +\fm{1424}{6561} f^3 -\left(\fm{17755}{243} 
\right. \right. \right. \nnb\\ &-& \left. \left. \left.
\fm{8992}{81} f +\fm{28}{9} f^2 \right) \overline{Q} \right]
-\fm{2247378701}{531441} - \fm{766276}{6561} \zeta_3 
+\left( \fm{195301451}{354294} - \fm{21056}{2187} \zeta_3\right) f 
-\left(\fm{1251734}{59049} 
\right. \right. \nnb\\ &+& \left. \left.
\fm{1540}{243} \zeta_3 \right) f^2 -\fm{1432}{6561} f^3 
+\left[ \fm{7375087}{3888} -\fm{40664}{27} \zeta_3 
       -\left(\fm{62269}{324} + \fm{142}{27} \zeta_3 \right) f
      -\fm{74}{81} f^2 \right] \overline{Q} \right\},\nnb\\[2mm]
Z_{68} &=&  -\fm{\al^2}{\ep} \left( \fm{1906}{81} - \fm{55}{27} f \right) 
+ \fm{\al^3}{\ep} \left[
\fm{1}{\ep} \left( \fm{873038}{2187} -\fm{9530}{243} f +\fm{695}{486} f^2 \right)
-\fm{726064}{729} -\fm{410713}{17496} f -\fm{6031}{2916} f^2 \right]
\nnb\\ &+& 
\fm{\al^4}{\ep} \left[ 
\fm{1}{\ep^2} \left( -\fm{131785021}{19683} +\fm{448189}{486} f
                     -\fm{380411}{8748} f^2 +\fm{80}{81} f^3 \right) 
+ \fm{1}{\ep} \left( \fm{184494545}{8748} -\fm{940982911}{629856} f
                    +\fm{3173417}{104976} f^2 
\right. \right. \nnb\\ &-& \left. \left.
\fm{2821}{972} f^3 \right) -\fm{97818607289}{5668704} -\fm{4146212}{2187} \zeta_3 
+\left( \fm{3023834561}{3779136} + \fm{309493}{1458} \zeta_3 \right) f
-\left( \fm{3278851}{78732} -\fm{175}{324} \zeta_3 \right) f^2 
\right. \nnb\\ &-& \left. 
\fm{997}{4374} f^3 \right].
\eea
\mathindent1cm
The $n$-loop $1/\ep^n$ poles vanish in the above renormalization
constants, which is a consequence of the lack of one-loop contributions
to $Z_{i7}$ and $Z_{i8}$.

The operators $X_i$ (\ref{x7x8}) differ by a multiplicative factor of
$1/\al$ from the bare versions of $Q_7$ and $Q_8$. We have used $X_i$
in the actual renormalization constant calculation. Each $n$-loop term
is then of order $\al^n$, which makes the simple matrix relation
between renormalization constants and anomalous dimensions applicable.
Such a relation up to three-loops has been given in Appendix~B of
Ref.~\cite{Chetyrkin:1997fm}. Extending that result to four loops in
the same notation (but with $\kappa^{0n}=0$) yields
\mathindent0cm
\bea 
\hat{\gamma} &=& 2 \hat{a}^{11} g^2 
+ g^4 \left[ 4 \hat{a}^{12} -\! 2 \hat{a}^{01}\hat{a}^{11}  -\! 2 \hat{a}^{11}\hat{a}^{01}  
           -\! 4 \kappa^{11} \hat{a}^{01} \right] 
+ g^6 \left[ 6 \hat{a}^{13} 
            -4 \hat{a}^{12} \hat{a}^{01} 
            -2 \hat{a}^{01} \hat{a}^{12}
            -4 \hat{a}^{02} \hat{a}^{11} 
\right. \nonumber \\ &-& \left.
            2 \hat{a}^{11} \hat{a}^{02} 
            +2 \hat{a}^{01} \hat{a}^{11} \hat{a}^{01} 
            +2 \hat{a}^{11} (\hat{a}^{01})^2
            +2 (\hat{a}^{01})^2 \hat{a}^{11} 
            +4 \kappa^{11} (\hat{a}^{01})^2
            -8 \kappa^{11} \hat{a}^{02} 
            -8 \kappa^{12} \hat{a}^{01} \right] 
\nonumber \\ &+& 
  g^8 \left[ 
             8 \hat{a}^{14} 
           - 6 \hat{a}^{13} \hat{a}^{01} 
           - 2 \hat{a}^{01} \hat{a}^{13} 
           - 6 \hat{a}^{03} \hat{a}^{11} 
           - 2 \hat{a}^{11} \hat{a}^{03} 
           - 4 \hat{a}^{12} \hat{a}^{02} 
           - 4 \hat{a}^{02} \hat{a}^{12} 
           + 4 \hat{a}^{02} \hat{a}^{01} \hat{a}^{11} 
\right. \nonumber \\ &+& \left. 
             4 \hat{a}^{02} \hat{a}^{11} \hat{a}^{01} 
         +\! 2 \hat{a}^{01} \hat{a}^{12} \hat{a}^{01} 
           + 2 \hat{a}^{11} \hat{a}^{02} \hat{a}^{01} 
           + 2 \hat{a}^{01} \hat{a}^{02} \hat{a}^{11} 
           + 2 \hat{a}^{11} \hat{a}^{01} \hat{a}^{02} 
           + 2 \hat{a}^{01} \hat{a}^{11} \hat{a}^{02} 
           + 4 \hat{a}^{12} (\hat{a}^{01})^2 
\right. \nonumber \\ &+& \left. 
             2 (\hat{a}^{01})^2 \hat{a}^{12} 
         -\! 2 \hat{a}^{01} \hat{a}^{11} (\hat{a}^{01})^2 
         -\! 2 (\hat{a}^{01})^2 \hat{a}^{11} \hat{a}^{01} 
         -\! 2 \hat{a}^{11} (\hat{a}^{01})^3 
         -\! 2 (\hat{a}^{01})^3 \hat{a}^{11} 
        -\! 12 \hat{a}^{03} \kappa^{11} 
        -\! 12 \hat{a}^{01} \kappa^{13}
\right. \nonumber \\ &-& \left. 
            16 \hat{a}^{02} \kappa^{12} 
           + 8 \hat{a}^{02} \hat{a}^{01} \kappa^{11} 
           + 4 \hat{a}^{01} \hat{a}^{02} \kappa^{11} 
           + 8 (\hat{a}^{01})^2 \kappa^{12} 
           - 4 (\hat{a}^{01})^3 \kappa^{11} \right] + {\cal O}(g^{10}). \label{gamma}
\eea
\mathindent1cm

After calculating the ADM $\hat{\gamma}$ from Eq.~(\ref{gamma}), we
multiply the dipole operators by $\al$, which brings us to the
normalization of $Q_7$ and $Q_8$ as in Eq.~(\ref{physical}). Next, we
pass from $\hat{\gamma}$ to $\hat{\gamma}^{\rm eff}$, which amounts to
performing a simple linear transformation\footnote{
See Eq.~(38) of Ref.~\cite{Buras:1994xp}.}
that stems from Eq.~(\ref{ceff}). 

The obtained complete expressions for $\hat{\gamma}^{(n)\rm eff}$
($n=0,1,2$) in the $b \to s \gamma$ case ($f=5$,
$\overline{Q}=\f{1}{3}$, $Q_u=\f{2}{3}$) are as follows:
\mathindent0cm 
\be \label{gamma0} \hat{\gamma}^{(0)\rm eff}
= \left[
\begin{array}{cccccccc}
\vspace{0.2cm}
-4 & \f{8}{3} &       0     &   -\f{2}{9} &      0    &     0     & -\f{208}{243} &  \f{173}{162} \\[2mm] 
12 &     0    &       0     &    \f{4}{3} &      0    &     0     &   \f{416}{81} &    \f{70}{27} \\[2mm] 
 0 &     0    &       0     &  -\f{52}{3} &      0    &     2     &  -\f{176}{81} &    \f{14}{27} \\[2mm] 
 0 &     0    &  -\f{40}{9} & -\f{100}{9} &  \f{4}{9} &  \f{5}{6} & -\f{152}{243} & -\f{587}{162} \\[2mm] 
 0 &     0    &       0     & -\f{256}{3} &      0    &    20     & -\f{6272}{81} &  \f{6596}{27} \\[2mm] 
 0 &     0    & -\f{256}{9} &   \f{56}{9} & \f{40}{9} & -\f{2}{3} & \f{4624}{243} &  \f{4772}{81} \\[2mm] 
 0 &     0    &       0     &       0     &      0    &     0     &     \f{32}{3} &        0      \\[2mm] 
 0 &     0    &       0     &       0     &      0    &     0     &    -\f{32}{9} &     \f{28}{3} \\[2mm]
\end{array} \right] \ee
\be \label{gamma1}
\hat{\gamma}^{(1)\rm eff} = \left[
\begin{array}{cccccccc}
\vspace{0.2cm}
-\f{355}{9} & -\f{502}{27} &  -\f{1412}{243} &  -\f{1369}{243} &    \f{134}{243} &   -\f{35}{162} &     -\f{818}{243} &     \f{3779}{324} \\[2mm] 
 -\f{35}{3} &   -\f{28}{3} &    -\f{416}{81} &    \f{1280}{81} &      \f{56}{81} &     \f{35}{27} &       \f{508}{81} &     \f{1841}{108} \\[2mm] 
     0      &        0     &   -\f{4468}{81} &  -\f{31469}{81} &     \f{400}{81} &  \f{3373}{108} &    \f{22348}{243} &     \f{10178}{81} \\[2mm] 
     0      &        0     &  -\f{8158}{243} & -\f{59399}{243} &    \f{269}{486} & \f{12899}{648} &   -\f{17584}{243} &  -\f{172471}{648} \\[2mm] 
     0      &        0     & -\f{251680}{81} & -\f{128648}{81} &   \f{23836}{81} &   \f{6106}{27} &  \f{1183696}{729} &  \f{2901296}{243} \\[2mm] 
     0      &        0     &  \f{58640}{243} & -\f{26348}{243} & -\f{14324}{243} & -\f{2551}{162} & \f{2480344}{2187} & -\f{3296257}{729} \\[2mm] 
     0      &        0     &         0       &         0       &         0       &        0       &      \f{4688}{27} &          0        \\[2mm]  
     0      &        0     &         0       &         0       &         0       &        0       &     -\f{2192}{81} &     \f{4063}{27}  \\[2mm]
\end{array} \right]. \ee
\bea \label{gamma2}
\hat{\gamma}^{(2)\rm eff} \hspace{-3mm} &=& \hspace{-2mm} \left[
\begin{array}{cccccccc}
\vspace{0.2cm}
-\f{12773}{18} & \f{745}{9} &     \f{63187}{13122} &     -\f{981796}{6561} & -\f{202663}{52488} &    -\f{24973}{69984} &   \f{150994745}{1062882} &     -\f{138745277}{354294} \\[2mm]
   \f{1177}{2} &        306 &     \f{110477}{2187} &      \f{133529}{8748} &   -\f{42929}{8748} &    \f{354319}{11664} &    \f{138336202}{177147} &      -\f{83347037}{472392} \\[2mm]
             0 &          0 &   -\f{3572528}{2187} &   -\f{58158773}{8748} &   \f{552601}{4374} &   \f{6989171}{11664} &    -\f{58397866}{177147} &     \f{5093967523}{944784} \\[2mm]
             0 &          0 &   -\f{1651004}{6561} & -\f{155405353}{52488} & \f{1174159}{52488} &  \f{10278809}{34992} & -\f{5108749081}{2125764} &   -\f{9826683847}{5668704} \\[2mm]
             0 &          0 & -\f{147978032}{2187} &  -\f{168491372}{2187} & \f{11213042}{2187} &   \f{17850329}{2916} &   \f{5824017302}{177147} &    \f{24198694001}{118098} \\[2mm]
             0 &          0 &  \f{136797922}{6561} &  -\f{72614473}{13122} & -\f{9288181}{6561} & -\f{16664027}{17496} &   \f{3603565835}{531441} & -\f{122956397803}{1417176} \\[2mm]
             0 &          0 &                    0 &                     0 &                  0 &                    0 &           \f{179768}{81} &                          0 \\[2mm]
             0 &          0 &                    0 &                     0 &                  0 &                    0 &           -\f{23444}{81} &             \f{192137}{81} 
\end{array} \right]
\nnb\\[5mm] && \hspace{-7mm} +~\zeta_3~ \left[
\begin{array}{cccccccc}
\vspace{0.2cm}
\f{1472}{3} & -\f{4288}{9} &  -\f{1360}{81} &    -\f{776}{81} &    \f{124}{81} &   \f{100}{27} &   \f{1272596}{6561} &  -\f{2373229}{4374} \\[2mm]
      -2144 &         -224 &   \f{2720}{27} &   -\f{2768}{27} &   -\f{248}{27} &   -\f{110}{9} &  -\f{2713672}{2187} &    -\f{682658}{729} \\[2mm]
          0 &            0 &   -\f{608}{27} &   \f{61424}{27} &   -\f{496}{27} &  -\f{2821}{9} &   \f{3236560}{2187} &   -\f{1051828}{729} \\[2mm]
          0 &            0 &  \f{88720}{81} &   \f{54272}{81} &  -\f{9274}{81} & -\f{3100}{27} &   \f{2007886}{6561} &   \f{4858889}{2187} \\[2mm]
          0 &            0 &  \f{87040}{27} &  \f{324416}{27} & -\f{13984}{27} & -\f{31420}{9} & \f{112180720}{2187} & -\f{119487244}{729} \\[2mm]
          0 &            0 & \f{721408}{81} & -\f{166432}{81} & -\f{95032}{81} & -\f{7552}{27} &  \f{15361912}{6561} & -\f{71430598}{2187} \\[2mm]
          0 &            0 &              0 &               0 &              0 &             0 &      -\f{21056}{27} &                   0 \\[2mm]
          0 &            0 &              0 &               0 &              0 &             0 &       \f{22976}{81} & -\f{48304}{27} 
\end{array} \right].~ \eea
\mathindent1cm
For arbitrary values of $f$, $\overline{Q}$ and $Q_u$, the blocks $B^{(n)}$ ($n=0,1,2$) read
\mathindent0cm
\bea
B^{(0)} &=& \left[ \begin{array}{ccc}
\f{8}{243} - \f{4}{3} Q_u                       && \f{173}{162}\\[2mm]
-\f{16}{81} + 8 Q_u                             && \f{70}{27}\\[2mm]
-\f{176}{81}                                    && \f{14}{27}\\[2mm]
\f{88}{243} - \f{16}{81} f                      && \f{74}{81} - \f{49}{54} f\\[2mm]
-\f{6272}{81}                                   && \f{1736}{27} + 36 f\\[2mm]
\f{3136}{243} - \f{160}{81} f + 48 \overline{Q} && \f{2372}{81} + \f{160}{27} f
\end{array} \right], \label{B0}\\[3mm]
B^{(1)} &=& \left[ \begin{array}{ccc}
\f{12614}{2187} - \f{64}{2187} f - \f{374}{27} Q_u + \f{2}{27} f Q_u && 
\f{65867}{5832} + \f{431}{5832} f\\[2mm]
-\f{2332}{729} + \f{128}{729} f + \f{136}{9} Q_u - \f{4}{9} f Q_u && 
\f{10577}{486} - \f{917}{972} f\\[2mm]
\f{97876}{729} - \f{4352}{729} f - \f{112}{3} \overline{Q} && 
\f{42524}{243} - \f{2398}{243} f\\[2mm]
-\f{70376}{2187} - \f{15788}{2187} f + \f{32}{729} f^2 - \f{140}{9} \overline{Q} && 
-\f{159718}{729} - \f{39719}{5832} f - \f{253}{486} f^2\\[2mm]
\f{1764752}{729} - \f{65408}{729} f - \f{3136}{3} \overline{Q} && 
\f{2281576}{243} + \f{140954}{243} f - 14 f^2\\[2mm]
\f{4193840}{2187} - \f{324128}{2187} f + \f{896}{729} f^2 - \f{1136}{9} \overline{Q} - \f{56}{3} f \overline{Q} && 
-\f{3031517}{729} - \f{15431}{1458} f - \f{6031}{486} f^2
\end{array} \right], \label{B1}\\[3mm]
B^{(2)} &=&  \left[ \begin{array}{l}
 \f{77506102}{531441}-\f{875374}{177147}f+\f{560}{19683}f^2-\f{9731}{162}Q_u+\f{11045}{729}f Q_u+\f{316}{729}f^2 Q_u+\f{3695}{486}\overline{Q}\\[2mm] 
-\f{15463055}{177147}+\f{242204}{59049}f-\f{1120}{6561}f^2+\f{55748}{27}Q_u-\f{33970}{243}f Q_u-\f{632}{243}f^2 Q_u-\f{3695}{81}\overline{Q}\\[2mm] 
 \f{102439553}{177147}-\f{12273398}{59049}f+\f{5824}{6561}f^2+\f{26639}{81}\overline{Q}-\f{8}{27}f \overline{Q}\\[2mm] 
-\f{2493414077}{1062882}-\f{9901031}{354294}f+\f{243872}{59049}f^2-\f{1184}{6561}f^3-\f{49993}{972}\overline{Q}+\f{305}{27}f \overline{Q}\\[2mm] 
 \f{8808397748}{177147}-\f{174839456}{59049}f+\f{1600}{729}f^2-\f{669694}{81}\overline{Q}+\f{10672}{27}f \overline{Q}\\[2mm] 
 \f{7684242746}{531441}-\f{351775414}{177147}f-\f{479776}{59049}f^2-\f{11456}{6561}f^3+\f{3950201}{243}\overline{Q}
                                                                 -\f{130538}{81}f \overline{Q}-\f{592}{81}f^2 \overline{Q}
\end{array} \right.\nnb\\[3mm]
&&  \left. \begin{array}{ll} \hspace{57mm}
&-\f{421272953}{1417176}-\f{8210077}{472392}f-\f{1955}{6561}f^2\\[2mm]
& \f{98548513}{472392}-\f{5615165}{78732}f-\f{2489}{2187}f^2\\[2mm]
& \f{3205172129}{472392}-\f{108963529}{314928}f+\f{58903}{4374}f^2\\[2mm]
&-\f{6678822461}{2834352}+\f{127999025}{1889568}f+\f{1699073}{157464}f^2+\f{505}{4374}f^3\\[2mm]
& \f{29013624461}{118098}-\f{64260772}{19683}f-\f{230962}{243}f^2-\f{148}{27}f^3\\[2mm]
&-\f{72810260309}{708588}+\f{2545824851}{472392}f-\f{33778271}{78732}f^2-\f{3988}{2187}f^3\\[2mm]
\end{array} \right]\nnb\\[3mm]
&+&  \zeta_3 \left[ \begin{array}{ll}
-\f{112216}{6561}+\f{728}{729}f+\f{25508}{81}Q_u-\f{64}{81}f Q_u-\f{100}{27}\overline{Q}&
-\f{953042}{2187}-\f{10381}{486}f\\[2mm]
 \f{365696}{2187}-\f{1168}{243}f-\f{51232}{27}Q_u-\f{1024}{27}f Q_u+\f{200}{9}\overline{Q}& 
-\f{607103}{729}-\f{1679}{81}f\\[2mm] 
 \f{3508864}{2187}-\f{1904}{243}f-\f{1984}{9}\overline{Q}-\f{64}{9}f \overline{Q}& 
-\f{1597588}{729}+\f{13028}{81}f-\f{20}{9}f^2\\[2mm] 
-\f{1922264}{6561}+\f{308648}{2187}f-\f{1280}{243}f^2+\f{1010}{9}\overline{Q}-\f{200}{27}f \overline{Q}& 
 \f{2312684}{2187}+\f{128347}{729}f+\f{920}{81}f^2\\[2mm] 
 \f{123543040}{2187}-\f{207712}{243}f+\f{128}{27}f^2-\f{24880}{9}\overline{Q}-\f{640}{9}f \overline{Q}& 
-\f{69359224}{729}-\f{885356}{81}f-\f{5080}{9}f^2\\[2mm]  
 \f{7699264}{6561}+\f{2854976}{2187}f-\f{12320}{243}f^2-\f{108584}{9}\overline{Q}-\f{1136}{27}f \overline{Q}& 
-\f{61384768}{2187}-\f{685472}{729}f+\f{350}{81}f^2
\end{array} \right]\nnb\\ \label{B2}
\eea
\mathindent1cm

The $f$- and charge-dependence of the blocks $C^{(n)}$ for $n=0,1,2$
can be found in Eqs.~(9)--(11) of Ref.~\cite{Gorbahn:2005sa}.  The
blocks $A^{(n)}$ are charge-independent, while their $f$-dependence
is given in Eqs.~(36)--(38) of Ref.~\cite{Gorbahn:2004my}.

\newsection{Numerical Analysis \label{sec:numerics}}

Let ${\cal B}_{\rm\scs NNLO}$ denote the NNLO value of the ${\bar
B}\to X_s\gamma$ branching ratio, and the superscript
``$4L\!\!\to\!\!0$'' indicate that the four-loop matrix $B^{(2)}$ is
set to zero in the calculation of a given quantity.  Normalizing the
difference between ${\cal B}_{\rm\scs NNLO}$ and ${\cal B}_{\rm\scs
NNLO}^{\rm\scs 4L\to 0}$ to the LO result, one finds
\be \label{NNLOratio}
\f{ {\cal B}_{\rm\scs NNLO} -
    {\cal B}_{\rm\scs NNLO}^{\rm\scs 4L\to 0}}{{\cal B}_{\rm\scs LO}} ~~=~~
\left( \f{\alpha_s(\mu_b)}{\pi}\right)^{\!\!2} \;\; \f{ C_7^{(2)\rm eff}(\mu_b) - 
\left[C_7^{(2)\rm eff}(\mu_b)\right]^{\rm\scs 4L\to 0}}{8\, C_7^{(0)\rm eff}(\mu_b)},
\ee
where $\mu_b$ is the low-energy scale at which the RGEs evolution is
terminated, while $C_7^{(n)\rm eff}$ are defined by the perturbative
expansion
\be \label{Wexp}
C_7^{\rm eff}(\mu) = \sum_{n\geq 0} \al^n(\mu) C_7^{(n)\rm eff}(\mu).
\ee

To evaluate the r.h.s. of Eq.~(\ref{NNLOratio}), we need to know the
non-vanishing LO initial conditions for the Wilson coefficients at the
electroweak scale $\mu_0$. They read~\cite{Inami:1980fz}
\bea
C^{(0)}_2(\mu_0) &=& 1,\\                               
C^{(0)}_7(\mu_0) &=& \f{3 x^3-2 x^2}{4(x-1)^4}\ln x + \f{-8 x^3 - 5 x^2 + 7 x}{24(x-1)^3},\\
C^{(0)}_8(\mu_0) &=& \f{-3 x^2}{4(x-1)^4}\ln x + \f{-x^3 + 5 x^2 + 2 x}{8(x-1)^3},
\eea
where $x = m_t^2(\mu_0)/M_W^2$. Substituting them into the analytical
solution to the RGEs (see, e.g., Section~3.3 of
Ref.~\cite{Huber:2005ig}), one obtains
\be \label{r.etas4}
\f{ C_7^{(2)\rm eff}(\mu_b) - \left[C_7^{(2)\rm eff}(\mu_b)\right]^{\rm\scs 4L\to 0}}{8\, C_7^{(0)\rm eff}(\mu_b)} ~=~
\f{h_1^{(2)} \eta^{a_1+2} + h_2^{(2)} \eta^{a_2+2} + \sum_{i=3}^8 h_i^{(2)} \eta^{a_i}}{ 
\eta^{a_2} C_7^{(0)}(\mu_0) + \f{8}{3} \left(\eta^{a_1}-\eta^{a_2}\right) C_8^{(0)}(\mu_0) + \sum_{i=1}^8 h_i^{(0)} \eta^{a_i}},
\ee
where $\eta=\alpha_s(\mu_0)/\alpha_s(\mu_b)$. For the numerical
  evaluation of $\eta$, we always use the four-loop RGE for
  $\alpha_s(\mu)$. The numbers $a_i$ and $h_i^{(n)}$ are given in
Table~\ref{tab:magic}. They are found from the diagonalization of
$\hat{\gamma}^{(0)\rm eff}$~(\ref{gamma0}) as well as the elements of
$\hat{\gamma}^{(1)\rm eff}$~(\ref{gamma1}) and $\hat{\gamma}^{(2)\rm
  eff}$~(\ref{gamma2}). Although $\hat{\gamma}^{(0)\rm eff}$ can be
  diagonalized analytically, the decimal approximations presented in
  Table~\ref{tab:magic} are much more convenient to use. Note that
$\sum_{i=1}^8 h_i^{(n)} = 0$, which follows from the initial condition for
$C_7$ at $\eta=1$ and from the off-diagonal position of $B^{(n)}$ in
$\hat{\gamma}^{(n)\rm eff}$.
\begin{table}[t]
\begin{tabular}{|l|rrrrrrrr|}
\hline
$i         $&$   1      $&$   2      $&$  3      $&$    4      $&$  5      $&$  6      $&$  7      $&$  8      $\\[1mm]
\hline
$a_i       $&$\f{14}{23}$&$\f{16}{23}$&$\f{6}{23}$&$-\f{12}{23}$&$  0.4086 $&$ -0.4230 $&$ -0.8994 $&$  0.1456 $\\[1mm]
$h_i^{(0)} $&$   2.2996 $&$  -1.0880 $&$ -0.4286 $&$   -0.0714 $&$ -0.6494 $&$ -0.0380 $&$ -0.0185 $&$ -0.0057 $\\[1mm]
$h_i^{(1)} $&$   0.9075 $&$  -0.7550 $&$  0.6971 $&$    0.0166 $&$ -0.7345 $&$  0.1784 $&$ -0.2608 $&$ -0.0493 $\\[1mm]
$h_i^{(2)} $&$ -13.9117 $&$  11.4063 $&$  3.4986 $&$   -0.0619 $&$ -2.4429 $&$  2.1210 $&$ -0.7558 $&$  0.1464 $\\\hline
\end{tabular}
\caption{ The numbers $a_i$ and $h_i^{(n)}$ that occur in Eqs.~(\ref{r.etas4})
  and (\ref{r.etas3}).\label{tab:magic}}
\end{table}

\begin{figure}[h]
\begin{center}
\includegraphics[width=55mm,angle=0]{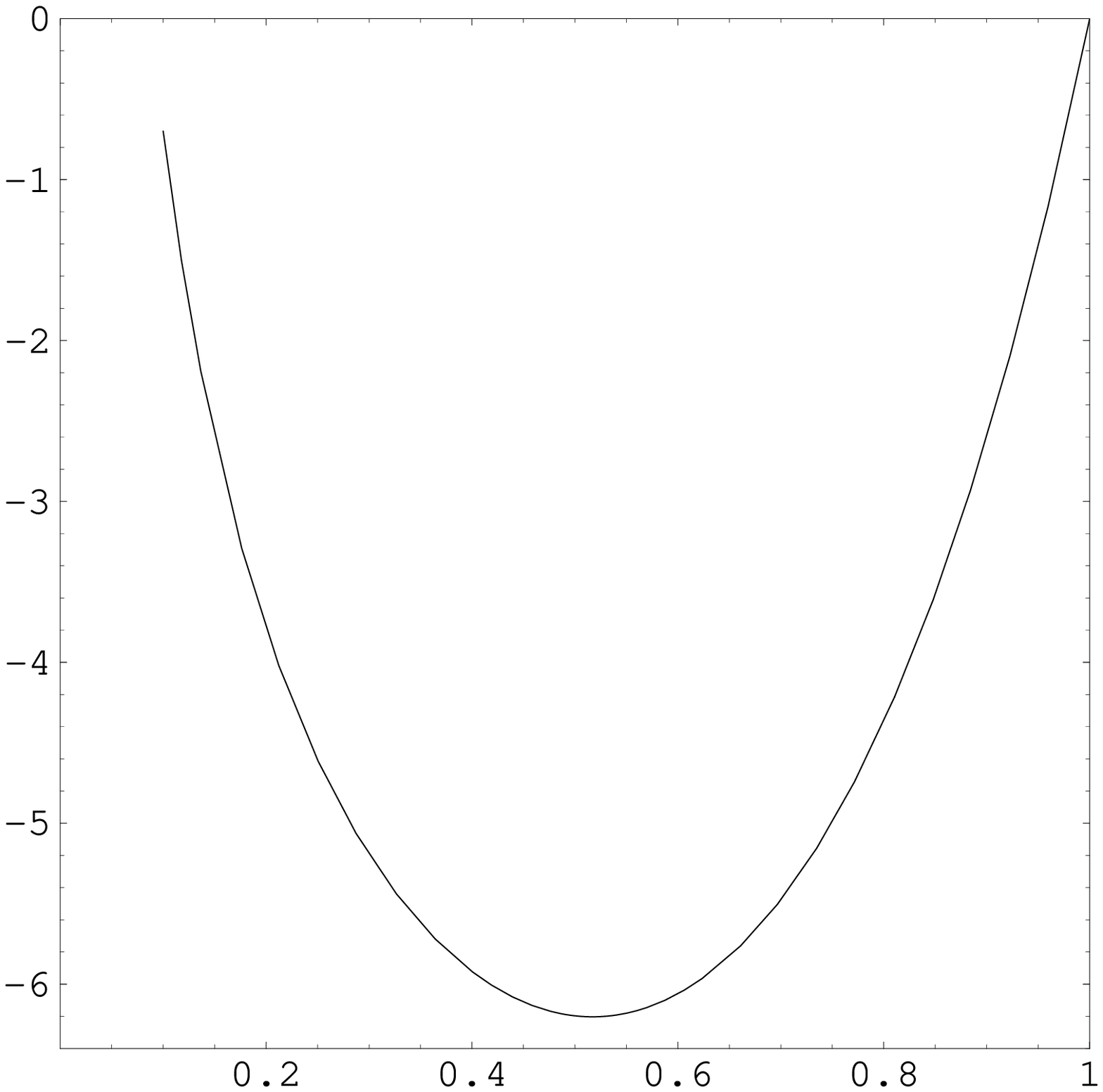}
\includegraphics[width=55mm,angle=0]{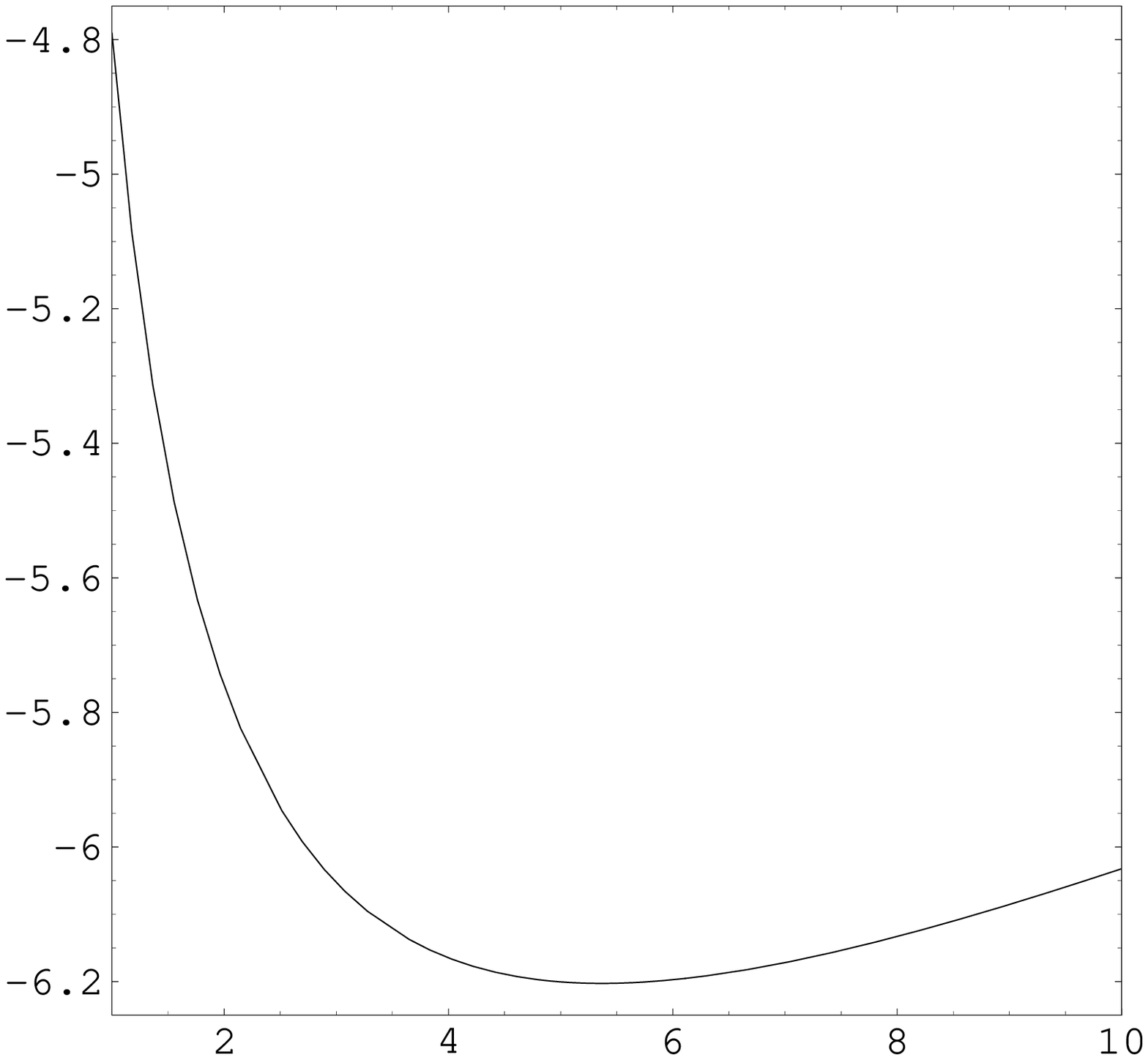}
\includegraphics[width=55mm,angle=0]{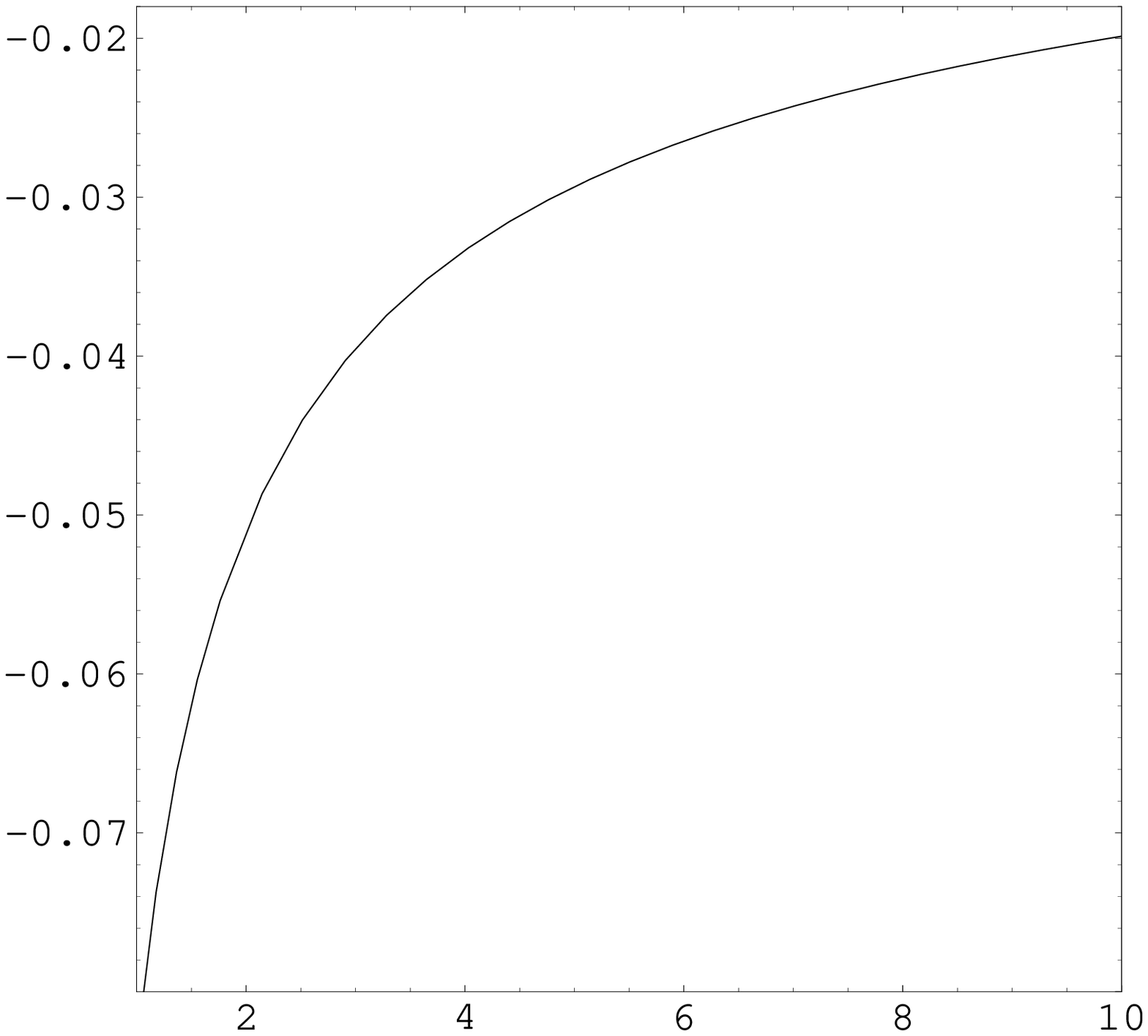}
\end{center}
\vspace*{-7mm}
\hspace*{27mm} $\eta$ \hspace{46mm} $\mu_b\;$[GeV] \hspace{39mm} $\mu_b\;$[GeV] \vspace{-3mm}
\begin{center}
\caption{\sf The r.h.s. of Eq.~(\ref{r.etas4}) as a function of $\eta$ (left plot) and $\mu_b$ (middle plot). The right plot
shows the relative NNLO correction (\ref{NNLOratio}) to the branching ratio as a function of $\mu_b$. \label{fig:numcor4}}
\end{center}
\vspace*{-12mm}
\end{figure}

The r.h.s. of Eq.~(\ref{r.etas4}) is shown in the left plot of
Fig.~\ref{fig:numcor4} as a function of $\eta$ for $x =
(162/80.4)^2$. The middle plot in the same figure presents this
quantity as a function of $\mu_b$ for $\mu_0 = 160\;$GeV and for the
remaining input parameters as listed in Appendix~A of
Ref.~\cite{Misiak:2006ab}. The $\mu_b$-dependence of the complete
correction (\ref{NNLOratio}) is shown for the same parameters in
the right plot of Fig.~\ref{fig:numcor4}. 

\begin{figure}[h]
\begin{center}
\includegraphics[width=55mm,angle=0]{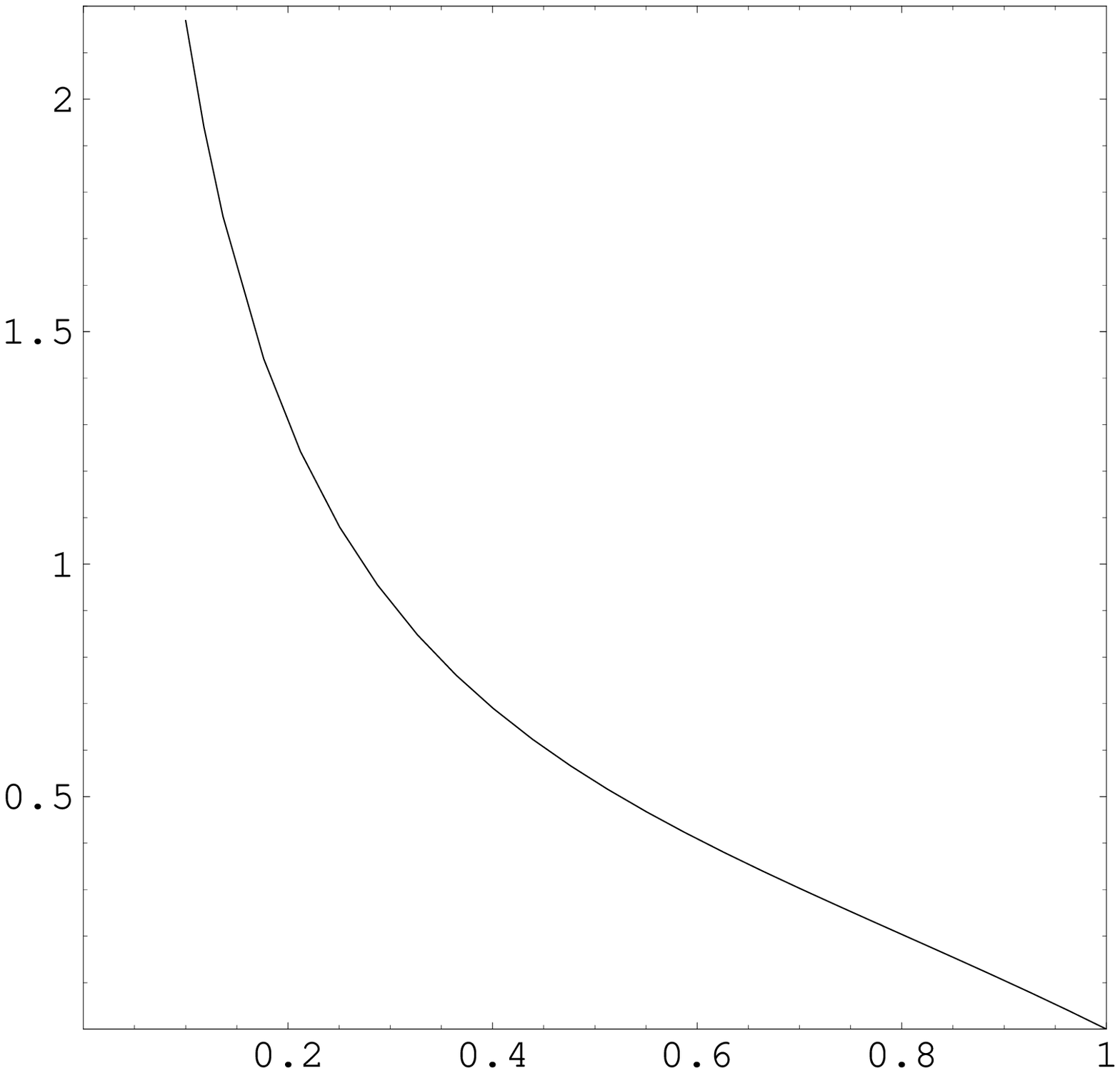}
\includegraphics[width=55mm,angle=0]{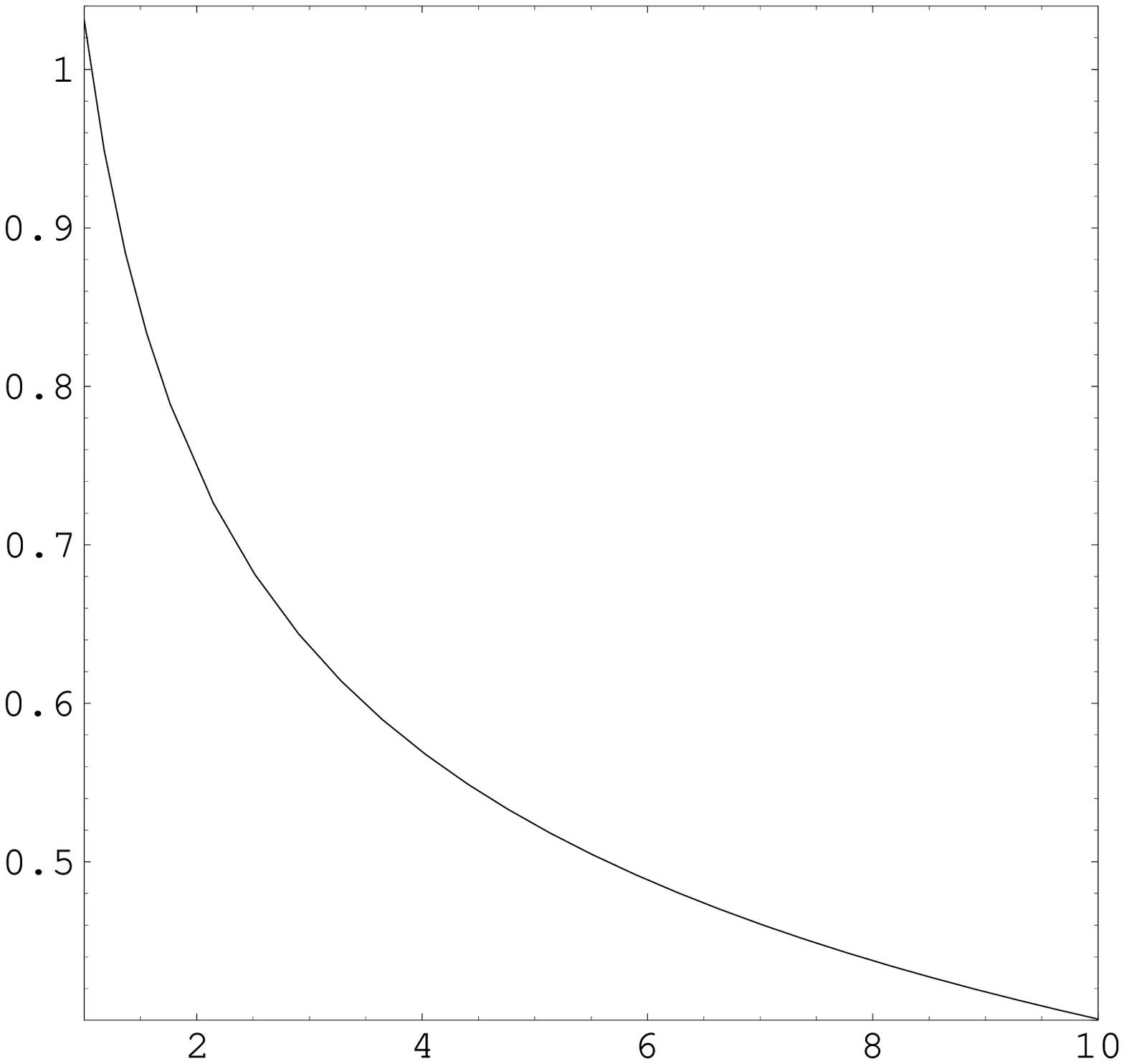}
\includegraphics[width=55mm,angle=0]{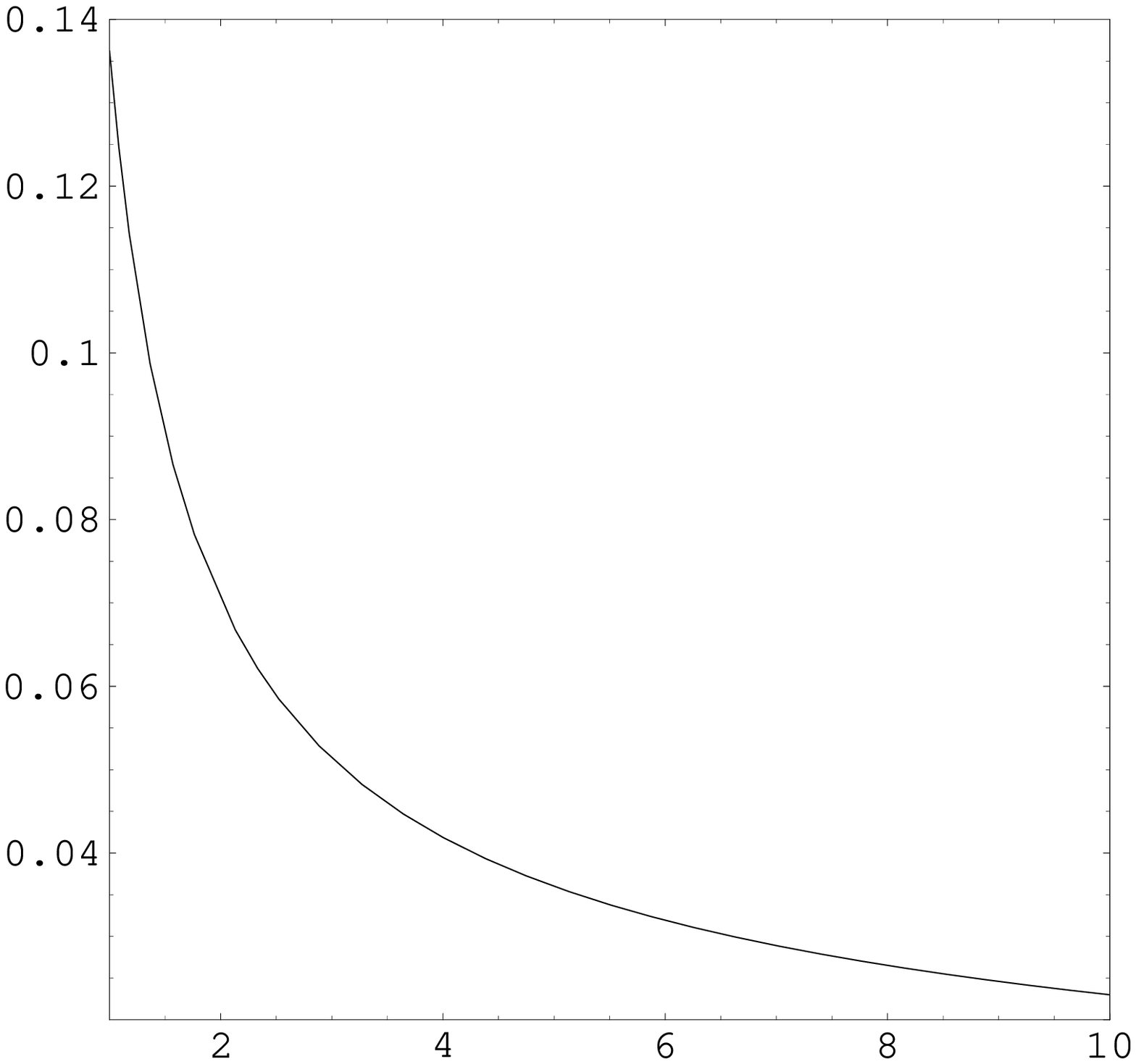}
\end{center}
\vspace*{-7mm}
\hspace*{27mm} $\eta$ \hspace{46mm} $\mu_b\;$[GeV] \hspace{39mm} $\mu_b\;$[GeV] \vspace{-3mm}
\begin{center}
\caption{\sf The r.h.s. of Eq.~(\ref{r.etas3}) as a function of $\eta$ (left plot) and $\mu_b$ (middle plot). The right plot
shows the relative NLO correction (\ref{NLOratio}) to the branching ratio as a function of $\mu_b$. \label{fig:numcor3}}
\end{center}
\vspace*{-5mm}
\end{figure}

It is interesting to compare the four-loop effect to the analogous
three-loop one at the Next-to-Leading Order (NLO). Let the superscript
``$3L\!\to\!0$'' indicate neglecting the three-loop matrix $B^{(1)}$.
Then
\be \label{NLOratio}
\f{ {\cal B}_{\rm\scs NLO} -
    {\cal B}_{\rm\scs NLO}^{\rm\scs 3L\to 0}}{{\cal B}_{\rm\scs LO}} ~~=~~
\f{\alpha_s(\mu_b)}{\pi} \;\; 
\f{ C_7^{(1)\rm eff}(\mu_b) - \left[C_7^{(1)\rm eff}(\mu_b)\right]^{\rm\scs 3L\to 0}}{2\, C_7^{(0)\rm eff}(\mu_b)},
\ee
and
\be \label{r.etas3}
\f{ C_7^{(1)\rm eff}(\mu_b) - \left[C_7^{(1)\rm eff}(\mu_b)\right]^{\rm\scs 3L\to 0}}{2\, C_7^{(0)\rm eff}(\mu_b)} ~=~
\f{h_1^{(1)} \eta^{a_1+1} + h_2^{(1)} \eta^{a_2+1} + \sum_{i=3}^8 h_i^{(1)} \eta^{a_i}}{ 
\eta^{a_2} C_7^{(0)}(\mu_0) + \f{8}{3} \left(\eta^{a_1}-\eta^{a_2}\right) C_8^{(0)}(\mu_0) + \sum_{i=1}^8 h_i^{(0)} \eta^{a_i}}.
\ee
The ratios (\ref{NLOratio}) and (\ref{r.etas3}) are shown in
Fig.~\ref{fig:numcor3} as functions of $\eta$ and $\mu_b$.

Comparing the right plots in Figs.~\ref{fig:numcor4} and \ref{fig:numcor3},
one can see that the two corrections are similar in magnitude. The suppression
of the NNLO one by an additional factor of~ $\alpha_s(\mu_b)/\pi$~ tends to
get compensated by larger values of $h_i^{(2)}$ as compared to $h_i^{(1)}$.
However, as the remaining plots show, the ratios (\ref{r.etas4}) and
(\ref{r.etas3}) do not contradict the naive expectation of being ``numbers of
order unity''. Thus, there seems to be no deep reason behind the observed
relative smallness of the NLO effect and the relatively large value of the
NNLO one. It should be remembered that the other NLO corrections exceed~ 
$\sim 30\%$ while the other NNLO ones stay within~ $\sim 10\%$ \cite{Misiak:2006zs}.

It is evident from Fig.~\ref{fig:numcor4} that the $\mu_b$-dependence of the NNLO
correction (\ref{NNLOratio}) for $\mu_b \in [1,10]\,$GeV originates 
almost entirely from the overall factor $\alpha_s^2(\mu_b)$.  The $\pm 3\%$
higher-order uncertainty in ${\cal B}_{\rm\scs NNLO}$ that was estimated in
Refs.~\cite{Misiak:2006zs,Misiak:2006ab} takes this $\mu_b$-dependence into
account. The central value of ${\cal B}_{\rm\scs NNLO}$ was calculated there
for $\mu_b = 2.5\,$GeV.  At that scale, our present correction amounts to
around~ $-4.4\%$. It is worth noting that ${\cal B}_{\rm\scs NNLO}^{\rm\scs
  4L\to 0}$ calculated along the formulae of Ref.~\cite{Misiak:2006ab} 
turns out to be much less $\mu_b$-dependent than the more complete
${\cal B}_{\rm\scs NNLO}$.

The actual numerical calculation in Refs.~\cite{Misiak:2006zs,Misiak:2006ab}
included contributions from $B^{(2)}_{i7}$ but not from the very recently
found $B^{(2)}_{i8}$. This fact is practically irrelevant for the above
discussion.  The numerical effect of $B^{(2)}_{i8}$ on ${\cal B}_{\rm\scs
  NNLO}$ is about 10 times smaller than the one of $B^{(2)}_{i7}$, and has the
same sign. The small contribution of $B^{(2)}_{i8}$ will be taken into
account together with other similar ones in a future upgrade of the
phenomenological NNLO analysis.

\newsection{Summary \label{sec:summary}}

We have evaluated the complete four-loop anomalous dimension matrix $B^{(2)}$
that is necessary for determining the effective FCNC couplings~ $\bar
qq'\gamma$~ and~ $\bar qq'g$~ at the NNLO in QCD. The results are presented in
a form that applies to any external flavour case, with an arbitrary number of
non-decoupled quarks. Adding an essential contribution to the NNLO QCD
analysis ${\bar B}\to X_s\gamma$~ has been the main purpose of our
calculation.  The obtained~ ${\cal O}(\alpha_s^2(\mu_b))$~ correction to the
branching ratio of this decay amounts to around~ $-2.9$\%~ for $\mu_b=5\,$GeV,
and~ $-4.4$\%~ for $\mu_b=2.5\,$GeV.

\newsection{Acknowledgments}

The work of M.C. was supported by the Sofja Kovalevskaja Award of the
Alexander von Humboldt Foundation sponsored by the German Federal Ministry of
Education and Research. U.H. was supported by the Swiss National Foundation.
M.M. acknowledges support by the Polish Committee for Scientific Research
under the grant 2~P03B~078~26, and from the EU Contract MRTN-CT-2006-035482,
FLAVIAnet. Our calculations made extensive use of 
the DESY-Zeuthen Grid Engine computer cluster,
the Fermilab General-Purpose Computing Farms~\cite{Albert:2003vv},
as well as the Z-box computer at the University of Z\"urich \cite{zbox}.

\begin{table}[t]
\begin{center}
\begin{tabular}{|c|c|c|c|}\hline
$k$     & \multicolumn{3}{c|}{$l$}    \\[1mm]\cline{2-4}
        & $n=0$   & $n=1$     & $n=2$ \\[1mm]\hline
1,2     & 1,2     &   2       & none  \\[1mm]
3,4,5,6 & 1,2,4   &   2       & none  \\[1mm]
7       & all     & 1,2,4,7,8 & 2,7,8 \\[1mm]
8       & 1,2,4,8 &   2,8     & none  \\\hline
\end{tabular}
\caption{\sf Indices $k$ and $l$ of the relevant $U^{(n)}_{kl}$
for $n=0,1,2$. \label{tab:indices}}
\end{center}
\end{table}

\newappendix{Appendix A}
\def\theequation{A.\arabic{equation}}

This appendix contains solutions to the RGEs for all the Wilson coefficients
that matter for $b\to s \gamma$ at the NNLO before including higher-order
electroweak corrections. A generic solution to the RGEs (\ref{RGEs})
reads
\be \label{RGEsol}
\vec{C}^{\rm eff}(\mu_b) = \hat{U}(\mu_b,\mu_0)\vec{C}^{\rm eff}(\mu_0).
\ee
The block-triangular structure of the matrix $\hat{U}$ corresponds to that of
$\left(\hat{\gamma}^{\rm eff}\right)^T$, taking into account that the blocks
$A^{(n)}$ and $C^{(n)}$ in Eq.~(\ref{blocks}) have a block-triangular form,
too. In particular, $U_{87}=0$, and $U_{kl}=0$ for $k\leq 2$ and $l>2$.

Using the perturbative expansions for $C_i^{\rm eff}(\mu)$ (\ref{Wexp}) and
\be
\hat{U}(\mu_b,\mu_0) = \sum_{n\geq 0} \al^n(\mu_0) \hat{U}^{(n)},
\ee
one easily finds that
\bea
\vec{C}^{(0)\rm eff}(\mu_b) &=& \hat{U}^{(0)} \vec{C}^{(0)\rm eff}(\mu_0),\\[2mm]
\vec{C}^{(1)\rm eff}(\mu_b) &=& \eta \left[ 
\hat{U}^{(0)} \vec{C}^{(1)\rm eff}(\mu_0)+
\hat{U}^{(1)} \vec{C}^{(0)\rm eff}(\mu_0)\right],\\[2mm]
\vec{C}^{(2)\rm eff}(\mu_b) &=& \eta^2 \left[ 
\hat{U}^{(0)} \vec{C}^{(2)\rm eff}(\mu_0)+
\hat{U}^{(1)} \vec{C}^{(1)\rm eff}(\mu_0)+
\hat{U}^{(2)} \vec{C}^{(0)\rm eff}(\mu_0)\right].
\eea
The matrices $\hat{U}^{(n)}$ are functions of $\eta$ only
\be \label{Usol}
U^{(n)}_{kl} = \sum_{j=0}^n \sum_{i=1}^8 
m^{(nj)}_{kli} \eta^{a_i-j}.
\ee
The powers $a_i$ have been given in Table~\ref{tab:magic}. 
Only a relatively small set of non-vanishing $U^{(n)}_{kl}$
is relevant for $b \to s \gamma$ at the NNLO, so long as
\bea
C_l^{(0)\rm eff}(\mu_0) &=& 0, \mbox{~~~~for~} l=1,3,4,5,6,\\
C_l^{(1)\rm eff}(\mu_0) &=& 0, \mbox{~~~~for~} l=3,5,6,
\eea
which is the case in the SM. The corresponding values of the indices $k$ and
$l$ for $n=0,1,2$ are collected in Table~\ref{tab:indices}. The ``magic numbers''
$m^{(nj)}_{kli}$ that occur in the expressions (\ref{Usol}) for these
$U^{(n)}_{kl}$ are given in Tables~\ref{tab:mag0}--\ref{tab:mag2}.

As far as $C_i^{(n)\rm eff}(\mu_0)$ are concerned, their SM values for
$i=1,\ldots,6$ can be found in Section~2 of Ref.~\cite{Bobeth:1999mk}. The
coefficients $C_i^{Q(n)}$ from that paper combine with a relative minus sign
to our $C_i^{(n)}$, i.e. $C_i^{(n)} = C_i^{t(n)}\! - C_i^{c(n)}$.  For
$i=7,8$, one can find the corresponding $C_i^{Q(n)}$ in Section~6 of
Ref.~\cite{Misiak:2004ew}. They combine to our $C_i^{(n)}$ according to the
relation $C_i^{(n)} = C_i^{t(n+1)}\! - C_i^{c(n+1)}$. The upper index gets
shifted because $Q_7$ and $Q_8$ in Ref.~\cite{Misiak:2004ew} were normalized
as $X_7$ and $X_8$ in Eq.~(\ref{x7x8}) here. Once $C_i^{(n)}(\mu_0)$ are
evaluated, Eq.~(\ref{ceff}) should be applied to obtain 
$C_i^{(n)\rm eff}(\mu_0)$.

\begin{table}[t]
\begin{center}
\begin{tabular}{|r|rrrrrrrr|}\hline
 $i$&1&2&3&4&5&6&7&8\\\hline
 $m^{(00)}_{11i}$&0&0&0.3333&0.6667&0&0&0&0\\[1mm]
 $m^{(00)}_{12i}$&0&0&1&$-$1&0&0&0&0\\[1mm]
 $m^{(00)}_{21i}$&0&0&0.2222&$-$0.2222&0&0&0&0\\[1mm]
 $m^{(00)}_{22i}$&0&0&0.6667&0.3333&0&0&0&0\\[1mm]
 $m^{(00)}_{31i}$&0&0&0.0106&0.0247&$-$0.0129&$-$0.0497&0.0092&0.0182\\[1mm]
 $m^{(00)}_{32i}$&0&0&0.0317&$-$0.0370&$-$0.0659&0.0595&$-$0.0218&0.0335\\[1mm]
 $m^{(00)}_{34i}$&0&0&0&0&$-$0.1933&0.1579&0.1428&$-$0.1074\\[1mm]
 $m^{(00)}_{41i}$&0&0&0.0159&$-$0.0741&0.0046&0.0144&0.0562&$-$0.0171\\[1mm]
 $m^{(00)}_{42i}$&0&0&0.0476&0.1111&0.0237&$-$0.0173&$-$0.1336&$-$0.0316\\[1mm]
 $m^{(00)}_{44i}$&0&0&0&0&0.0695&$-$0.0459&0.8752&0.1012\\[1mm]
 $m^{(00)}_{51i}$&0&0&$-$0.0026&$-$0.0062&0.0018&0.0083&$-$0.0004&$-$0.0009\\[1mm]
 $m^{(00)}_{52i}$&0&0&$-$0.0079&0.0093&0.0094&$-$0.0100&0.0010&$-$0.0017\\[1mm]
 $m^{(00)}_{54i}$&0&0&0&0&0.0274&$-$0.0264&$-$0.0064&0.0055\\[1mm]
 $m^{(00)}_{61i}$&0&0&$-$0.0040&0.0185&0.0021&$-$0.0136&$-$0.0043&0.0012\\[1mm]
 $m^{(00)}_{62i}$&0&0&$-$0.0119&$-$0.0278&0.0108&0.0163&0.0103&0.0023\\[1mm]
 $m^{(00)}_{64i}$&0&0&0&0&0.0317&0.0432&$-$0.0675&$-$0.0074\\[1mm]
 $m^{(00)}_{71i}$&0.5784&$-$0.3921&$-$0.1429&0.0476&$-$0.1275&0.0317&0.0078&$-$0.0031\\[1mm]
 $m^{(00)}_{72i}$&2.2996&$-$1.0880&$-$0.4286&$-$0.0714&$-$0.6494&$-$0.0380&$-$0.0185&$-$0.0057\\[1mm]
 $m^{(00)}_{73i}$&8.0780&$-$5.2777&0&0&$-$2.8536&0.1281&0.1495&$-$0.2244\\[1mm]
 $m^{(00)}_{74i}$&5.7064&$-$3.8412&0&0&$-$1.9043&$-$0.1008&0.1216&0.0183\\[1mm]
 $m^{(00)}_{75i}$&202.9010&$-$149.4668&0&0&$-$55.2813&2.6494&0.7191&$-$1.5213\\[1mm]
 $m^{(00)}_{76i}$&86.4618&$-$59.6604&0&0&$-$25.4430&$-$1.2894&0.0228&$-$0.0917\\[1mm]
 $m^{(00)}_{77i}$&0&1&0&0&0&0&0&0\\[1mm]
 $m^{(00)}_{78i}$&2.6667&$-$2.6667&0&0&0&0&0&0\\[1mm]
 $m^{(00)}_{81i}$&0.2169&0&0&0&$-$0.1793&$-$0.0730&0.0240&0.0113\\[1mm]
 $m^{(00)}_{82i}$&0.8623&0&0&0&$-$0.9135&0.0873&$-$0.0571&0.0209\\[1mm]
 $m^{(00)}_{84i}$&2.1399&0&0&0&$-$2.6788&0.2318&0.3741&$-$0.0670\\[1mm]
 $m^{(00)}_{88i}$&1&0&0&0&0&0&0&0\\\hline
\end{tabular}
\caption{\sf ``Magic numbers'' for the relevant $U^{(0)}_{kl}.$\label{tab:mag0}}
\end{center}
\end{table}

\begin{table}[t]
\begin{center}
\begin{tabular}{|r|rrrrrrrr|}\hline
 $i$&1&2&3&4&5&6&7&8\\\hline
 $m^{(10)}_{12i}$&0&0&$-$2.9606&$-$4.0951&0&0&0&0\\[1mm] 
 $m^{(11)}_{12i}$&0&0&5.9606&1.0951&0&0&0&0\\[1mm]
 $m^{(10)}_{22i}$&0&0&$-$1.9737&1.3650&0&0&0&0\\[1mm] 
 $m^{(11)}_{22i}$&0&0&1.9737&$-$1.3650&0&0&0&0\\[1mm]
 $m^{(10)}_{32i}$&0&0&$-$0.0940&$-$0.1517&$-$0.2327&0.2288&0.1455&$-$0.4760\\[1mm] 
 $m^{(11)}_{32i}$&0&0&$-$0.5409&1.6332&1.6406&$-$1.6702&$-$0.2576&$-$0.2250\\[1mm]
 $m^{(10)}_{42i}$&0&0&$-$0.1410&0.4550&0.0836&$-$0.0664&0.8919&0.4485\\[1mm] 
 $m^{(11)}_{42i}$&0&0&2.2203&2.0265&$-$4.1830&$-$0.7135&$-$1.8215&0.7996\\[1mm]
 $m^{(10)}_{52i}$&0&0&0.0235&0.0379&0.0330&$-$0.0383&$-$0.0066&0.0242\\[1mm] 
 $m^{(11)}_{52i}$&0&0&0.0400&$-$0.1861&$-$0.1669&0.1887&0.0201&0.0304\\[1mm]
 $m^{(10)}_{62i}$&0&0&0.0352&$-$0.1138&0.0382&0.0625&$-$0.0688&$-$0.0327\\[1mm] 
 $m^{(11)}_{62i}$&0&0&$-$0.2614&$-$0.1918&0.4197&0.0295&0.1474&$-$0.0640\\[1mm]
 $m^{(10)}_{71i}$&0.0021&$-$1.4498&0.8515&0.0521&0.6707&0.1220&$-$0.0578&0.0355\\[1mm] 
 $m^{(11)}_{71i}$&$-$4.3519&3.0646&1.5169&$-$0.5013&0.3934&$-$0.6245&0.2268&0.0496\\[1mm]
 $m^{(10)}_{72i}$&9.9372&$-$7.4878&1.2688&$-$0.2925&$-$2.2923&$-$0.1461&0.1239&0.0812\\[1mm] 
 $m^{(11)}_{72i}$&$-$17.3023&8.5027&4.5508&0.7519&2.0040&0.7476&$-$0.5385&0.0914\\[1mm]
 $m^{(10)}_{74i}$&$-$8.6840&8.5586&0&0&0.7579&0.4446&0.3093&0.4318\\[1mm] 
 $m^{(11)}_{74i}$&$-$42.9356&30.0198&0&0&5.8768&1.9845&3.5291&$-$0.2929\\[1mm]
 $m^{(10)}_{77i}$&0&7.8152&0&0&0&0&0&0\\[1mm] 
 $m^{(11)}_{77i}$&0&$-$7.8152&0&0&0&0&0&0\\[1mm]
 $m^{(10)}_{78i}$&17.9842&$-$18.7604&0&0&0&0&0&0\\[1mm] 
 $m^{(11)}_{78i}$&$-$20.0642&20.8404&0&0&0&0&0&0\\[1mm]
 $m^{(10)}_{82i}$&3.7264&0&0&0&$-$3.2247&0.3359&0.3812&$-$0.2968\\[1mm] 
 $m^{(11)}_{82i}$&$-$5.8157&0&1.4062&$-$3.9895&3.2850&3.6851&$-$0.1424&0.6492\\[1mm]
 $m^{(10)}_{88i}$&6.7441&0&0&0&0&0&0&0\\[1mm] 
 $m^{(11)}_{88i}$&$-$6.7441&0&0&0&0&0&0&0\\\hline
\end{tabular}
\caption{\sf ``Magic numbers'' for the relevant $U^{(1)}_{kl}.$\label{tab:mag1}}
\end{center}
\end{table}

\begin{table}[t]
\begin{center}
\begin{tabular}{|r|rrrrrrrr|}\hline
 $i$&1&2&3&4&5&6&7&8\\\hline
 $m^{(20)}_{72i}$&$-$212.4136&167.6577&5.7465&$-$3.7262&28.8574&$-$2.1262&2.2903&0.1462\\[1mm] 
 $m^{(21)}_{72i}$&$-$74.7681&58.5182&$-$13.4731&3.0791&7.0744&2.8757&3.5962&$-$1.2982\\[1mm] 
 $m^{(22)}_{72i}$&31.4443&$-$18.1165&23.2117&13.2771&$-$19.8699&4.0279&$-$8.6259&2.6149\\[1mm]
 $m^{(20)}_{77i}$&0&44.4252&0&0&0&0&0&0\\[1mm] 
 $m^{(21)}_{77i}$&0&$-$61.0768&0&0&0&0&0&0\\[1mm] 
 $m^{(22)}_{77i}$&0&16.6516&0&0&0&0&0&0\\[1mm]
 $m^{(20)}_{78i}$&15.4051&$-$18.7662&0&0&0&0&0&0\\[1mm] 
 $m^{(21)}_{78i}$&$-$135.3141&146.6159&0&0&0&0&0&0\\[1mm] 
 $m^{(22)}_{78i}$&36.4636&$-$44.4043&0&0&0&0&0&0\\\hline
\end{tabular}
\caption{\sf ``Magic numbers'' for the relevant $U^{(2)}_{kl}.$\label{tab:mag2}}
\end{center}
\end{table}

\setlength {\baselineskip}{0.2in}
 

\begin{thebibliography}{99}
%
\bibitem{Bertolini:1990if}
  S.~Bertolini, F.~Borzumati, A.~Masiero and G.~Ridolfi,
  Nucl.\ Phys.\ B {\bf 353} (1991) 591.
%
\bibitem{Cho:1993zb}
  P.~L.~Cho and M.~Misiak,
  Phys.\ Rev.\ D {\bf 49} (1994) 5894
  [hep-ph/9310332];\\
  K.~Fujikawa and A.~Yamada,
  Phys.\ Rev.\ D {\bf 49} (1994) 5890.
%
\bibitem{Misiak:2006zs}
  M.~Misiak {\it et al.},
  hep-ph/0609232, to be published in the Phys.\ Rev.\ Lett.
%
\bibitem{Misiak:2006ab}
  M.~Misiak and M.~Steinhauser,
  hep-ph/0609241, to be published in the Nucl.\ Phys.\ B.
%
\bibitem{Buchalla:1995vs}
  G.~Buchalla, A.~J.~Buras and M.~E.~Lautenbacher,
  Rev.\ Mod.\ Phys.\  {\bf 68} (1996) 1125
  [hep-ph/9512380].
%
\bibitem{Chetyrkin:1996vx}
K.G.~Chetyrkin, M.~Misiak and M.~M\"unz,
Phys.\ Lett.\ B {\bf 400} (1997) 206,
Phys.\ Lett.\ B {\bf 425} (1998) 414 (E)
[hep-ph/9612313].
%
\bibitem{Buras:1994xp}
A.J.~Buras, M.~Misiak, M.~M\"unz and S.~Pokorski,
Nucl.\ Phys.\ B {\bf 424} (1994) 374
[hep-ph/9311345].
%
\bibitem{Gambino:2003zm}
  P.~Gambino, M.~Gorbahn and U.~Haisch,
  Nucl.\ Phys.\ B {\bf 673} (2003) 238
  [hep-ph/0306079].
%
\bibitem{Gorbahn:2004my}
M.~Gorbahn and U.~Haisch,
Nucl.\ Phys.\ B {\bf 713} (2005) 291 
[hep-ph/0411071].
%
\bibitem{Gorbahn:2005sa}
M.~Gorbahn, U.~Haisch and M.~Misiak,
Phys.\ Rev.\ Lett.\  {\bf 95} (2005) 102004
[hep-ph/0504194].
%
\bibitem{Misiak:1994zw}
  M.~Misiak and M.~M\"unz,
  Phys.\ Lett.\ B {\bf 344} (1995) 308
  [hep-ph/9409454].
%
\bibitem{Chetyrkin:1997fm}
  K.~G.~Chetyrkin, M.~Misiak and M.~M\"unz,
  Nucl.\ Phys.\ B {\bf 518} (1998) 473
  [hep-ph/9711266].
%
\bibitem{vanRitbergen:1997va}
  T.~van Ritbergen, J.~A.~M.~Vermaseren and S.~A.~Larin,
  Phys.\ Lett.\ B {\bf 400} (1997) 379
  [hep-ph/9701390].
%
\bibitem{Czakon:2004bu}
  M.~Czakon,
  Nucl.\ Phys.\ B {\bf 710} (2005) 485
  [hep-ph/0411261].
%
\bibitem{Buras:1989xd}
A.J.~Buras and P.H.~Weisz,
Nucl.\ Phys.\ B {\bf 333} (1990) 66;\\
M.~J.~Dugan and B.~Grinstein, Phys.\ Lett.\ B {\bf 256} (1990) 239.
%
\bibitem{Bobeth:1999mk}
C.~Bobeth, M.~Misiak and J.~Urban,
Nucl.\ Phys.\ B {\bf 574} (2000) 291
[hep-ph/9910220].
%
\bibitem{Misiak:2004ew}
M.~Misiak and M.~Steinhauser,
Nucl.\ Phys.\ B {\bf 683} (2004) 277 
[hep-ph/0401041].
%
\bibitem{Huber:2005ig}
  T.~Huber, E.~Lunghi, M.~Misiak and D.~Wyler,
  Nucl.\ Phys.\ B {\bf 740} (2006) 105
  [hep-ph/0512066].
%
\bibitem{Inami:1980fz}
  T.~Inami and C.~S.~Lim,
  Prog.\ Theor.\ Phys.\  {\bf 65} (1981) 297,
  Prog.\ Theor.\ Phys.\  {\bf 65} (1981) 1772 (E).
%
\bibitem{Albert:2003vv}
M.~Albert {\it et al.},
FERMILAB-TM-2209.
%
\bibitem{zbox} {\tt http://krone.physik.unizh.ch/\~{}stadel/zbox/start}
%
\end{thebibliography}
\end{document}